\documentclass{article}

\usepackage{arxiv}

\usepackage[utf8]{inputenc} % allow utf-8 input
\usepackage[T1]{fontenc}    % use 8-bit T1 fonts
\usepackage{hyperref}       % hyperlinks
\usepackage{url}            % simple URL typesetting
\usepackage{booktabs}       % professional-quality tables
\usepackage{amsfonts}       % blackboard math symbols
\usepackage{nicefrac}       % compact symbols for 1/2, etc.
\usepackage{microtype}      % microtypography
\usepackage{lipsum}
\usepackage{graphicx,verbatim,float,subfigure,graphicx,color,siunitx}
\usepackage{amsmath}
\graphicspath{ {./images/} }

\usepackage{amssymb}
\usepackage{svg}
\usepackage{multirow}  %Adds multi-row in tables
\usepackage[export]{adjustbox}
\usepackage{soul}
\usepackage{textcomp}
\usepackage{multirow}
\usepackage{enumitem}

\title{Densification Mechanics of Polymeric Syntactic Foams}

\author{
 P. Prabhakar$^*$, H. Feng, S. P Subramaniyan\\
  Dept. of Mechanical Engineering \\
  University of Wisconsin-Madison
   \vspace{0.05in} \\
  \texttt{$^*$pavana.prabhakar@wisc.edu}
\And
 M. Doddamani\\
  Dept. of Mechanical Engineering \\
  National Institute of Technology Karnataka\\
  Surathkal, India
}

\begin{document}
\maketitle
%\nodate{ }
 
\newcommand{\SPS}[1]{\textcolor{red}{\bf{Sabari: #1}}}
%\newcommand{\VD}[1]{\textcolor{blue}{\bf{ #1}}}
%\newcommand{\pavana}[1]{\textcolor{blue}{\bf{ #1}}}
%----------------------------------------------------------------------------------------
%	TITLE SECTION
%----------------------------------------------------------------------------------------

\begin{abstract}
In this paper, a fundamental understanding of the densification mechanics of polymeric syntactic foams under compressive loading is established. These syntactic foams are closed cell composite foams with thin-walled microballoons dispersed in a matrix (resin) whose closed cell structure provides excellent mechanical properties, like high strength and low density. There are several parameters that can contribute towards their mechanical properties, including, microballoon volume fraction, microballoon wall thickness, bonding between the microballoons and the matrix, and the crushing strength of microballoons. Conducting purely experimental testing by varying these parameters can be very time sensitive and expensive. Also, identification of densification mechanics is challenging using experiments only. Higher densification stress and energy are favorable properties under foam compression or crushing. Hence, the influence of key structural and material parameters associated with syntactic foams that dictate the mechanics of densification is studied here by implementing micromechanics based computational models and multiple linear regression analysis. Specifically, specific densification stresses and energy, which are densification stresses and energy normalized by weight, are evaluated which are more relevant for a wide variety of weight saving applications. Microballoon crushing strength and volume fraction are identified as the parameters that have the higher influence on densification stress and energy, and their specific counterparts, whereas the interfacial bonding has the least impact. In addition, designing aspects of syntactic foams with specified overall density are discussed by mapping microballoon volume fraction and wall thickness. The regression model allows for establishing wall thicknesses and corresponding volume fractions that result in higher densification properties for a specified overall foam density.
\end{abstract}

% keywords can be removed

%----------------------------------------------------------------------------------------
%	ARTICLE CONTENTS
\section{Introduction}\label{intro}
Polymeric syntactic foams are closed cell composite foams with thin-walled hollow microparticles or microspheres called `microballoons' dispersed in a polymer matrix resin. The closed cell structure provides excellent mechanical properties, like high strength and low density, in addition to lower moisture absorption as compared to open cell foams \cite{Gupta2002,Gupta2004Plastics,GladyszChawla2006,Choqueuse2008,Devi2007,BANHART2001}. Few widely known applications of these syntactic foams are in components for boat decks, ribs, hulls, and floatation modules for offshore structures \cite{Gupta2014}. In addition, they are also used in deep sea applications like remote operated vehicles, submarines, and underwater pipelines. In such applications, these syntactic foams are often subjected to severe compressive loading, which requires in-depth understanding of their densification mechanics for effective designing of structures.

%Few potential applications of syntactic foams is as core structures in sandwich composites that could be in building facades, bridge decks, and other civil infrastructure.

Few common microballoons used in syntactic foams include glass microballoons (GMB) and fly ash cenospheres. Past researchers have investigated the behavior of syntactic foams with engineering glass (Sodalime-borosilicate) microballoons \cite{Zhu2012,YUNG2009,JAYAVARDHAN2017,Poveda2013} and with fly ash cenospheres \cite{Gupta2001,Satapathy2011,Qiao2011,BharathKumar2016,Kumar2016,Barkoula2008} as the filler material. Extensive studies on the mechanical behavior of syntactic foams have been performed by previous researchers exploring their suitability for a wide range of applications \cite{Zhu2012,Gupta2010,Gladysz2006,Shunmugasamy2013}. Different types of tests have been performed on syntactic foams, such as three-point bending tests in flexure \cite{Maharsia2006,Tagliavia2010,Tagliavia2012} and short beam shear tests \cite{Kishore2005} to determine their response under such types of loading. In the work by Gupta et al. \cite{Gupta2004}, it was shown that the compressive strength and modulus of syntactic foams increased as the internal radius of cenospheres was reduced, that is as the wall thickness increased, while holding all other parameters fixed. Previous work by the current authors \cite{GARCIA2018265,Shahapurkar2018} investigated the behavior of cenosphere reinforced syntactic foams in compression and flexure over a range of temperatures. In Shahapurkar et al. \cite{Shahapurkar2018}, it was observed that the compressive modulus of cenosphere/epoxy syntactic foams increased and strength decreased with increasing cenosphere volume fraction. Additional data analysis showed that the failure strains decreased as the cenosphere volume fraction increased. In Jayavardhan and Doddamani \cite{JAYAVARDHAN2018}, the quasi-static compressive response of compression molded glass microballoon/High-Density Polyethylene (HDPE) syntactic foam was determined. They reported that the compressive modulus increased, but the yield strength, densification stress and overall energy absorption reduced with increasing microballoon volume fraction. However, the weight normalized values of yield strength, densification stress and overall energy absorption increased with increasing microballoon volume fraction.

In order to design syntactic foams with preferred compressive response, it is critical that the densification mechanics in terms of overall energy absorption and densification stress is well understood. There are several parameters that can influence the energy absorption and densification stress during compression of syntactic foams, such as microballoon volume fraction, microballoon wall thickness, strength of the microballoons, bonding between the microballoon and the matrix, among others. The influence of these key parameters needs to be elucidated for designing syntactic foams with better energy absorption capacity. Performing a purely experimental parametric study is very time consuming and expensive, especially when these parameters are considered independently while maintaining other parameters fixed. To that end, computational models can be used for isolating the impact of a single parameter such that its influence can be understood in detail. 

A majority of prior computational micromechanical modeling work \mbox{\cite{CAROLAN2020,NIAN2016,Shan2015,Cho2017}} have focused on predicting the tensile properties of syntactic foams, and seldom work on modeling the compressive behavior at the microscale. Work by Carolan et al. \mbox{\cite{CAROLAN2020}} combined numerical and experimental approach to predict the tensile moduli and failure strengths of syntactic foams. In addition, they were able to connect the scatter in fracture strength observed experimentally to their microstructure.
Nian et al. \mbox{\cite{NIAN2016}} studied the influence of interface properties and void content of syntactic foams on their tensile behavior using axisymmetric unit cell models. The interface debonding and matrix cracking were captured using cohesive zone modeling and Extended Finite Element Method, respectively. They showed that the tensile strength reduces with increasing particle volume fraction and increases with increasing particle wall thickness. Shan et al. \mbox{\cite{Shan2015}} investigated the influence of volume fraction of hollow particles, wall thickness, and the interface properties on the tensile strength of syntactic foams using a 3D micromechanical finite element model. They demonstrated that the interfacial strength between particles and matrix can improve the tensile strength of the syntactic foams. In the current paper, 2D micromechanical computational models are developed to elucidate the densification mechanics of GMB/HDPE syntactic foams under compressive loading by considering the key parameters mentioned above.

The current computational study is motivated by a prior experimental work on the compressive response of GMB/HDPE syntactic foams presented by Jayavardhan and Doddamani \cite{JAYAVARDHAN2018}. In that paper, the influence of microballoon volume fraction, microballoon wall thickness, and applied strain rate on the compressive behavior of GMB/HDPE syntactic foams was investigated. However, in our current paper, we have not considered the effect of strain rate and is beyond the scope of our work presented here. Further, in Jayavardhan and Doddamani \cite{JAYAVARDHAN2018}, GMBs of different densities at varying microballoon volume fractions were fabricated and tested under compressive loading. Neat HDPE samples were also tested to form a baseline response for comparison against GMB/HDPE foam samples. Primarily, energy absorption  and densification stresses reported in that paper \cite{JAYAVARDHAN2018} are of interest for the current study, and are most relevant for densification mechanics. They reported that in general, the densification stresses and energy absorption of GMB/HDPE foams increased with wall thickness (which relates to their crushing strength), however, reduced with increasing microballoon volume fraction. Further, thick walled particles that had higher crushing strengths survived more in number compared to thin walled microballoons. These densification mechanisms directly contribute to the densification stresses and corresponding absorption energy, and need to be understood for effectively designing these syntactic foams. The independent effect of wall thickness or radius ratio, crushing strength, and microballoon volume fraction in addition to other key parameters like interface bonding requires a very large number of experimentation, and was beyond the scope of that paper \cite{JAYAVARDHAN2018}. Towards that, in the current paper, our micromechanical models will be used for performing a parametric study to elucidate the importance of different parameters associated with syntactic foams. These models and modeling approach, although developed for GMB/HDPE syntactic foams, can be used for other syntactic foams.

The focus of our current paper is to establish the densification mechanics of syntactic foams, that is, to elucidate how crushing of foams occur and establish their influence on energy absorption and corresponding densification stresses. Parametric study performed using micromechanical models developed here will elucidate the influence of microballoon volume fraction, microballoon wall thickness, bonding between the microballoons and the matrix, and the strength of microballoons on the densification mechanics of syntactic foams. Densification mechanics in terms of microballoon crushing and collective impact on energy absorption is compared qualitatively with experimental observations in Jayavardhan and Doddamani \cite{JAYAVARDHAN2018}. Finally, parameters that have higher influence on the energy absorption and densification stresses are identified using a multiple linear regression modeling. In addition, we have shown through a case study how these trained models can be used for designing syntactic foams for a specified foam density and with improved densification behaviors.
 
\section{Computational Modeling}\label{compModel}
Micromechanical computational model developed here consists of a finite element method based micromechanics model with GMBs embedded in HDPE matrix. A description of the computational model including the boundary value problem and the micromechanics domain is presented in sections \ref{bvp} and \ref{microDomain}, respectively. Description of the materials, which includes the constitutive material models of GMBs and HDPE matrix, and GMB/HDPE interface definition is provided in section \ref{matDescription}. Finally, the parametric space considered in this paper is presented in section \ref{parametricSpace}.

\subsection{Boundary Value Problem}\label{bvp} 
\begin{figure}[H]
\centering
	\includegraphics[width=2.in]{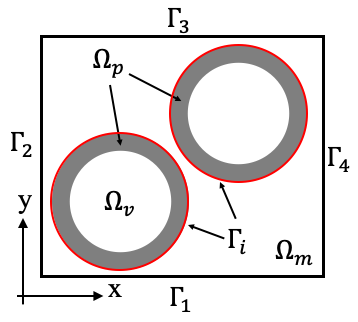}
	\caption{\small \sl Schematic of the 2D micromechanical syntactic foam model}\label{fig:modelDomain}
\end{figure}

2D micromechanical domain with randomly distributed microballoons in a matrix region is considered as the computational modeling domain. A schematic of the 2D micromechanical domain is shown in Figure~\ref{fig:modelDomain}. Here, $\Omega_{m}$ is the matrix region, $\Omega_{p}$ is the microballoon wall region, $\Omega_{v}$ is the hollow region inside the microballoons, such that the total domain $\Omega = \Omega_{p} \cup \Omega_{v} \cup \Omega_{m}$. $\Gamma_{1-4}$ are the external boundaries of the micromechanics domain, such that $\Gamma = \Gamma_1 \cup \Gamma_2 \cup \Gamma_3 \cup \Gamma_4$. $\Gamma_{i}$ are the interfaces between each microballoon and the matrix region. The matrix volume fraction, which in this 2D model is the matrix area fraction, is given by V$_{m}=\frac{\Omega_{m}}{\Omega}$. Correspondingly, the volume fraction of microballoons, including that of the particle wall and void, in a syntactic foam composite is V$_{mb}=1-$V$_{m} = \frac{\Omega_{p}+\Omega_{v}}{\Omega}$.

In the current paper, we consider a full Lagrangian formulation to allow finite deformations in the equilibrium equations for the domain. The boundary value problem (BVP) is formulated in the reference geometry. At time $t=0$, let the initial reference domain be $\Omega_0$, and the deformed domain at any time $t$ be $\Omega_t$. Let the coordinate of a point in the reference and the deformed domains be $x$ and $\hat{x}$, which gives us a displacement of $\boldsymbol{u}=\boldsymbol{\hat{x}}-\boldsymbol{x}$. Consequently, the deformation gradient ${\bf F}$, Cauchy-Green strain ($\boldsymbol{C}$) and first Piola-Kirchoff stress ($\boldsymbol{P}$) tensors are defined as,
%\vspace{-0.3in}
\begin{equation}
\begin{split}
    \boldsymbol{F}= \frac{\partial \boldsymbol{\hat{x}}}{\partial \boldsymbol{x}} = 1 + \frac{\partial \boldsymbol{u}}{\partial \boldsymbol{x}}, \quad 
    \boldsymbol{C} = \boldsymbol{F^T}\boldsymbol{F}, \quad
    \boldsymbol{P} = \frac{\partial U}{\partial \boldsymbol{F}}
\end{split}
\end{equation}

\noindent where, $U$ is the strain energy density. The two material domains, $\Omega_{p}$ and $\Omega_{m}$, are defined using individual material descriptions, which is discussed later in section \ref{matDescription}.

The boundary value problem solved for determining the compressive stress-strain response of syntactic foams is given by:

\begin{equation}
\begin{split}
\nabla.{\boldsymbol{P}}+ {\boldsymbol{B}} = 0 \quad \textrm{in} \quad  \Omega \\
\boldsymbol{P}.\boldsymbol{N} = \hat{\boldsymbol{T}} \quad \textrm{on} \quad \Gamma_T \quad \textrm{and} \quad \boldsymbol{u} = \boldsymbol{\hat{u}} \quad \textrm{on} \quad \Gamma_u, \quad
\Gamma_T \cap \Gamma_u = \emptyset, \quad \Gamma_T \cup \Gamma_u = \Gamma
\end{split}
\end{equation}

\noindent where, $\boldsymbol{B}$ is the reference body force. $\hat{T}$ is the applied traction on boundary edges, $N$ is the unit vector normal to the boundary edges in the reference configuration. $\hat{u}$ is the prescribed displacement on boundary edge, $\Gamma$, details of which are given below. 

Compressive displacement $\Delta_{applied}$ is applied on $\Gamma_{4}$. $\Gamma_{1}$ is restricted in motion along the y-axis and $\Gamma_{2}$ is restricted in motion along the x-axis. Flat boundary condition is considered on $\Gamma_{3}$, such that, the boundary is free to deform in the x and y directions, however, the displacement in the y-axis is constant along that boundary. This condition is commonly used in micromechanics modeling of composites. The corresponding boundary conditions are:

\begin{equation}\label{appDisp}
\begin{split}
    \hat{u}_{x} = - \Delta_{applied} \quad \textrm{on} \quad \Gamma_{4} \\
 \hat{u}_{x} = 0 \quad \textrm{on} \quad \Gamma_{2} \\
 \hat{u}_{y} = 0 \quad \textrm{on} \quad \Gamma_{1} \\
 \hat{u}_{y} = \textrm{constant} \; \textrm{on} \; \Gamma_{3}
 \end{split}
\end{equation}

\subsection{Micromechanics Domain}\label{microDomain}

\begin{figure}[h!]
\centering
\subfigure[]{
\includegraphics[width=0.25\textwidth, height=0.25\textwidth]{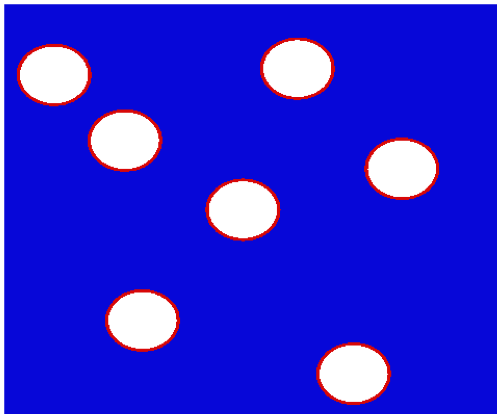}
}
%\centering
\subfigure[]{
\includegraphics[width=0.25\textwidth, height=0.25\textwidth]{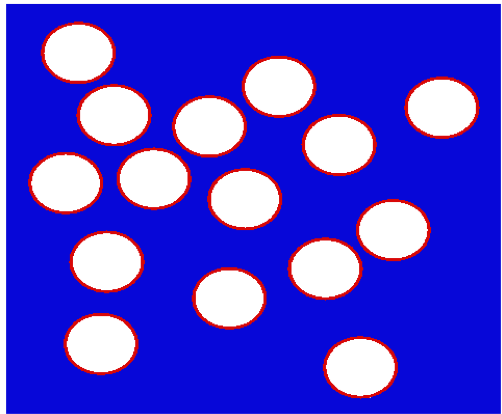}
} 
%\centering
\subfigure[]{
\includegraphics[width=0.25\textwidth, height=0.25\textwidth]{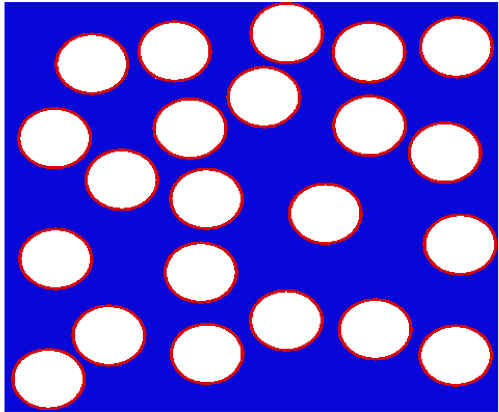}
}
%\centering
\subfigure[]{
\includegraphics[width=0.25\textwidth]{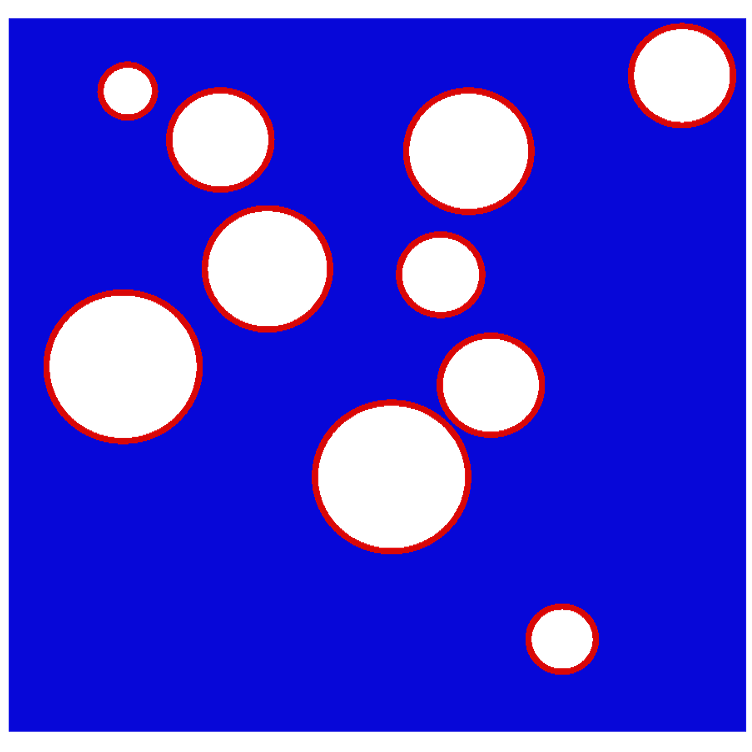}
}
%\centering
\subfigure[]{
\includegraphics[width=0.25\textwidth]{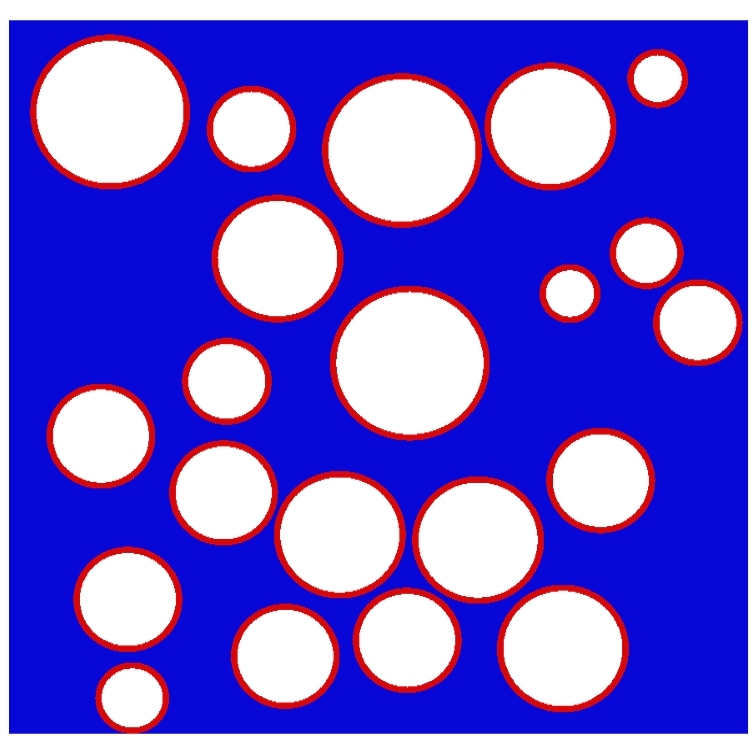}
} 
%\centering
\subfigure[]{
\includegraphics[width=0.25\textwidth]{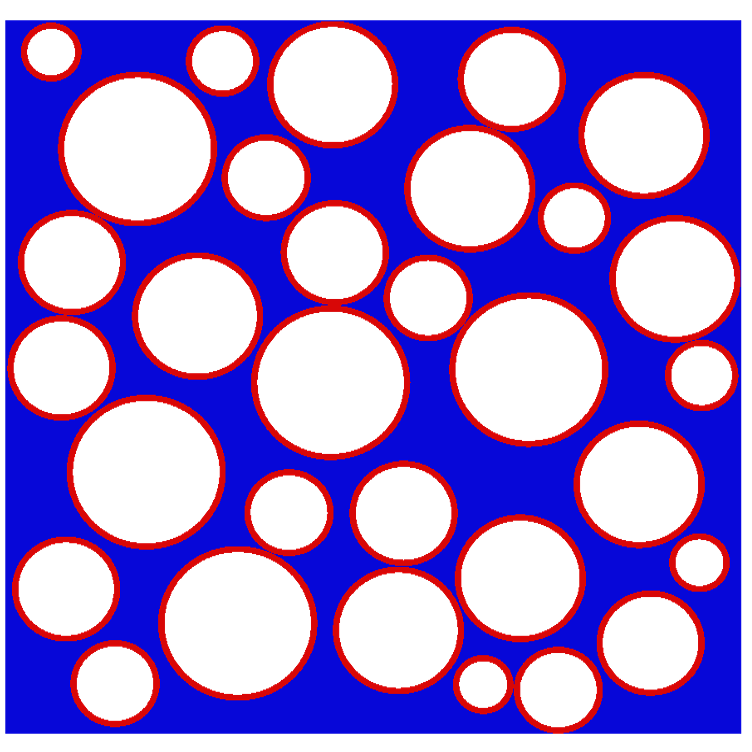}
}
%\centering
\subfigure[]{
\includegraphics[width=0.25\textwidth]{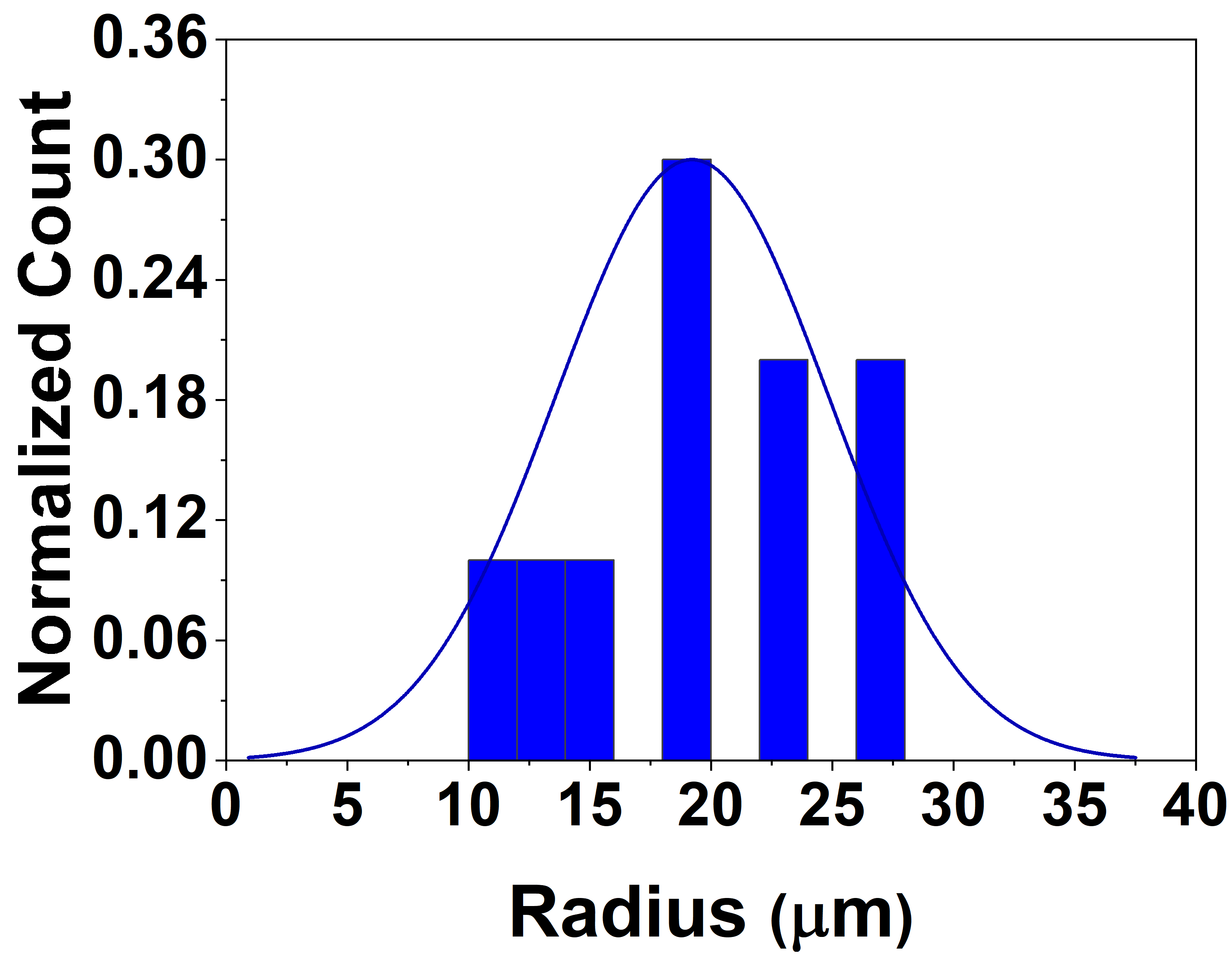}
}
%\centering
\subfigure[]{
\includegraphics[width=0.25\textwidth]{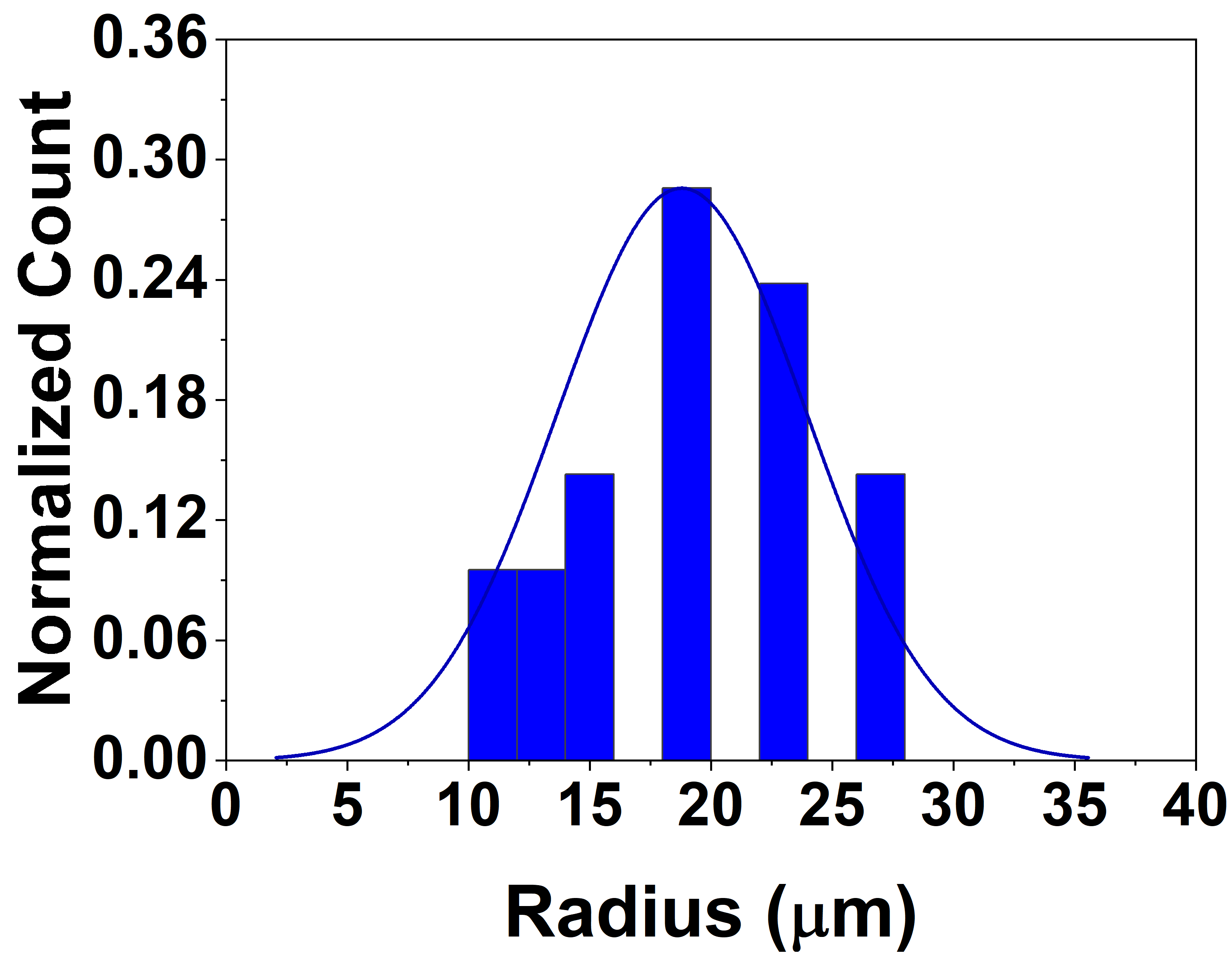}
} 
%\centering
\subfigure[]{
\includegraphics[width=0.25\textwidth]{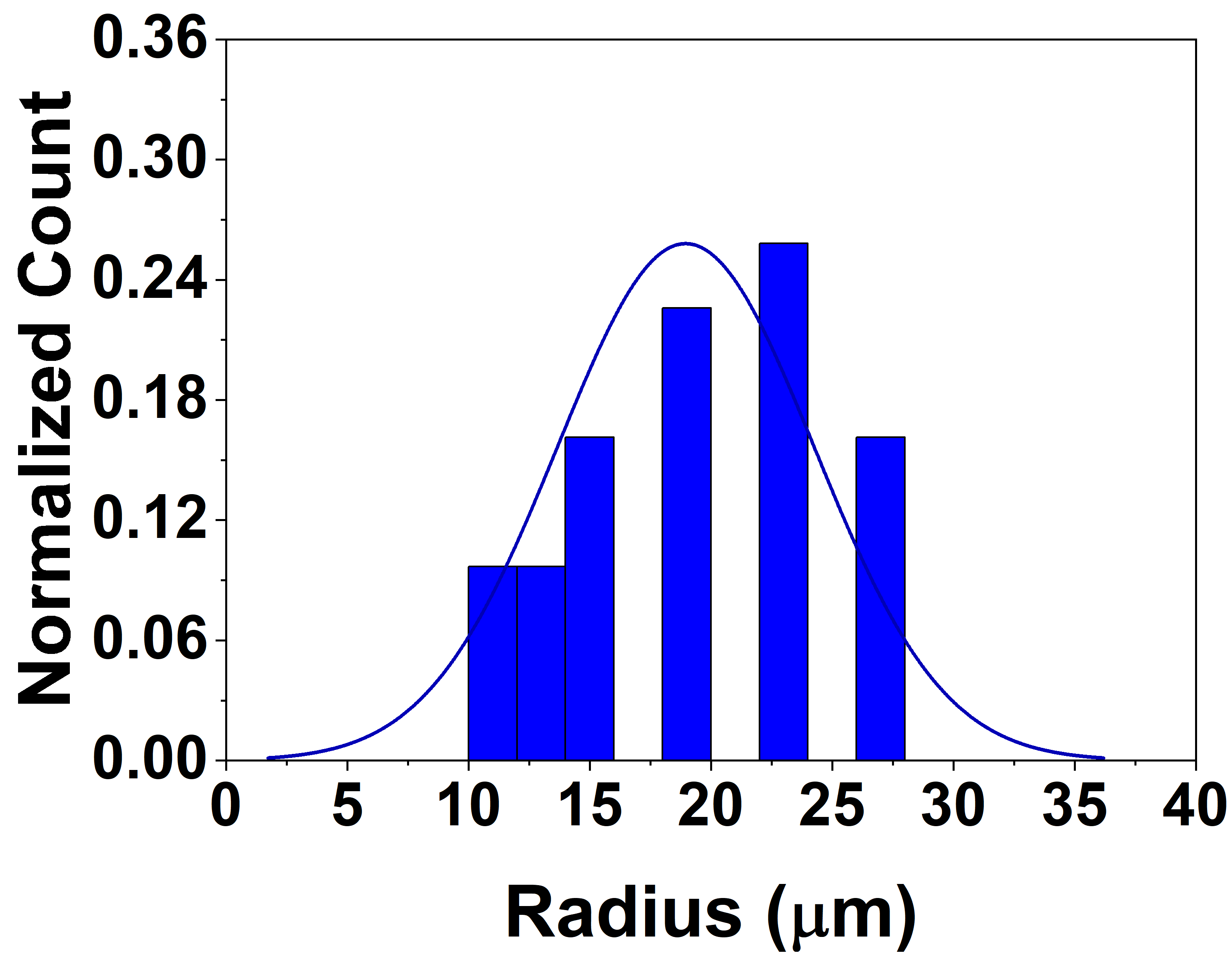}
}
\caption{Representative 2D micromechanical models of GMB/HDPE syntactic foams with different volume fractions of embedded microballoons: microballoons with uniform size - (a) 20\%, (b) 40\%, (c) 60\%;  microballoons with distributed size - (d) 20\%, (e) 40\%, (f) 60\%. Blue region is matrix, red region is GMB particles. Microballoon size distributions for each volume fraction - (g) 20\%, (h) 40\%, (i) 60\%} \label{fig:packing}
\end{figure}

Representative micromechanical domains with varying microballoon volume fraction (V$_{mb}$) maintained at 20\%, 40\% and 60\% by volume, and both uniform and distributed microballoon diameters, are shown in \mbox{Figure~\ref{fig:packing}}. A random sequential adsorption algorithm is used to generate the micromechanical models where particles are added sequentially in a domain without overlap until it achieves the desired particle volume fraction. The particle size is determined based on the probability density function of the overall material which is experimentally obtained. The mean and standard deviation of the particle radius \mbox{($\mu m$)} distribution are 19.22 $\pm$ 5.62, 18.8 $\pm$ 5.14, and 18.96 $\pm$ 5.29 for 20\%, 40\%, and 60\%, respectively, as compared to 20.19 $\pm$ 8.57 for the experimentally measured particle size distribution.

Model dimensions are maintained at 0.295 mm x 0.295 mm. This model size captures the microballoon distribution as well as provides results within 10\% of the converged solution. Hence, this model size strikes a good balance between convergence and computational cost. These models are implemented within ABAQUS \mbox{\cite{abaqus}} finite element method software with a mesh consisting of linear plane stress elements (CPS3 and CPS4). Since we have not considered strain softening material behavior in our models, there is no pathological mesh dependency. 
% Approximate element sizes are chosen to be 1 micron and 5 micron for the GMB and HDPE regions, respectively, to avoid element failure during meshing and convergence issues.
The parametric space explored in this study with regard to model geometry and material properties is described later in section~\ref{parametricSpace}. Note that four micromechanical models with randomly distributed microballoons for each volume fraction are simulated, but only one of each is shown in Figure~\ref{fig:packing}. Displacement boundary conditions applied on these models are given in Equations \ref{appDisp}. The corresponding reaction forces are determined after solving the boundary value problem described in the previous section, which are then used for establishing their compressive stress-strain responses.

\subsection{Constitutive Materials Description}\label{matDescription}

{\bf HDPE matrix}

HDPE manifests a non-linear behavior under compression as shown in \cite{JAYAVARDHAN2018}, which is represented here using a hyperelastic constitutive model. In particular, Ogden hyperelastic model \cite{ogden1972} is used here with third order fit (i.e. n=3) within ABAQUS \cite{abaqus}, a commercially available finite element analysis software. 

The strain energy potential of Ogden model in terms of principle stretches ($\lambda_1$,$\lambda_2$ and $\lambda_3$,) is given by:
\begin{equation}
\begin{split}
    % U := \sum_{i=1}^{n} \frac{2\mu_i}{\alpha_i^2} (\bar{\lambda}_1^{\alpha_i}+\bar{\lambda}_2^{\alpha_i}+\bar{\lambda}_3^{\alpha_i}-3) + \sum_{i=1}^{n} \frac{1}{D_i} (J_{el}-1)^{2i} \\
    U := \sum_{i=1}^{n} \frac{2\mu_i}{\alpha_i^2} ({\bar\lambda}_1^{\alpha_i}+{\bar\lambda}_2^{\alpha_i}+{\bar\lambda}_3^{\alpha_i}-3) 
    + \sum_{i=1}^N \frac{1}{D_i} (J_{el}-1)^{2i} \\
    \bar\lambda_i = J^{-\frac{1}{3}} \lambda_i \rightarrow \bar\lambda_1  \bar\lambda_2  \bar\lambda_3 = 1 
    %where, \bar{\lambda}_i = J^{-\frac{1}{3}} \lambda_i \rightarrow \bar{\lambda}_1 \bar{\lambda}_2 \bar{\lambda}_3 = 1
    \end{split}
\end{equation}

where, $J_{el}$ is the elastic volume ratio. Constants $\mu_i$ and $\alpha_i$ describe the shear behavior, and $D_i$ describes the compressibility of the material. $\lambda_j$, where j=1,2,3, are the principal stretches defined as the ratio of the deformed length ($l_j$) to the undeformed length ($L_i$) of a differential cubic volume element along the principal axes of a Cartesian coordinate system.
\begin{equation}
    \lambda_j = \frac{l_j}{L_j} \quad j \in [1,2,3]
\end{equation}

Principal stresses can be obtained by taking derivatives of the strain energy potential with respect to the principle stretches as:
\begin{equation}
    P_j = \frac{\partial U}{\partial \lambda_j} = J^{-\frac{1}{3}} \frac{\partial U}{\partial \bar\lambda_j}
\end{equation}

%For incompressible or almost incompressible materials, we have:
%\begin{equation}\label{incompress}
%    \lambda_1 \; . \; \lambda_2 \; . \; \lambda_3 = 1
%\end{equation}

If a uniaxial stress is applied on a differential cubic volume element along direction-1, the corresponding stretch $\lambda_1 = \lambda$, which is related to uniaxial strain $\epsilon$ as $ \lambda = \epsilon + 1 $. The stretches along the other two directions 2 and 3 are equal, i.e., $\lambda_2 = \lambda_3$. This gives $\bar\lambda_1 = \bar\lambda = J^{-\frac{1}{3}} \lambda$, and $\bar\lambda_2 = \bar\lambda_3 = \bar\lambda^{-\frac{1}{2}}$. Substituting these in the strain-energy equation, we have,
\begin{equation}
    U := \sum_{i=1}^{n} \frac{2\mu_i}{\alpha_i^2} ({\bar\lambda}^{\alpha_i}+2{\bar\lambda}^{-\frac{1}{2}\alpha_i}-3) 
\end{equation}

For deriving the stress-stretch relation for uniaxial compression/tension, we take the first derivative of the strain energy with respect to the uniaxial stretch ($\lambda$), which results in,
\begin{equation}
\begin{aligned}
%\begin{split}
\boldsymbol{P}_{1} &:= J^{-\frac{1}{3}} \frac{\partial U (\bar\lambda)}{\partial \bar\lambda}\\
    \boldsymbol{P}_1 &= J^{-\frac{1}{3}} \sum_{i=1}^{n} \frac{2\mu_i}{\alpha_i} ({\bar\lambda}^{\alpha_i-1}-{\bar\lambda}^{-\frac{1}{2}\alpha_i-1}) \quad \textrm{[Compressible]} \\
    \boldsymbol{P}_1 &= \sum_{i=1}^{n} \frac{2\mu_i}{\alpha_i} ({\lambda}^{\alpha_i-1}-{\lambda}^{-\frac{1}{2}\alpha_i-1}) \quad \textrm{[Incompressible]}
%    \end{split}
\end{aligned}
\end{equation}

The compressive stress-strain response of virgin HDPE is shown in Figure~\ref{fig:hdpeTest}, which is obtained from \cite{JAYAVARDHAN2018} and is used as an input to fit the Ogden hyperelastic model described above. The parameters obtained after fitting the Ogden model with a third order fit are shown in Table \ref{ogdenFit} and the corresponding nominal stress-strain response is shown in Figure~\ref{fig:hdpeTest}. This fitted model is used for representing the hyperelastic behavior of the matrix region (HDPE) in subsequent syntactic foam models with varying microballoon volume fractions. The Ogden hyperelastic constitutive model considered here is assumed to be isotropic with $\nu = 0.425$. Non-linear elastic response under large strain conditions is considered, and visco-elastic or stress-softening (damage) effects are ignored.
\begin{table}[H]
  \centering
  \begin{tabular}{cc}
    \begin{minipage}{.5\textwidth}
    \begin{figure}[H]
        \includegraphics[width=2.5in]{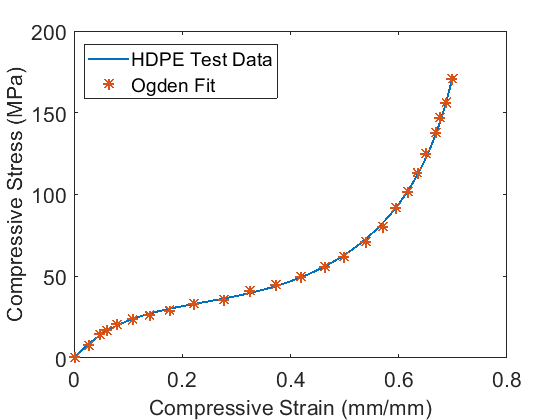}
        \caption{Compressive stress-strain response of HDPE}
        \label{fig:hdpeTest}
    \end{figure}
    \end{minipage} 
    &
    \begin{minipage}{.45\textwidth}
    \begin {table}[H]
	\caption {Parameters of the fitted Ogden model}
	\label{ogdenFit}
	\centering
	\begin{tabular}{|c|c|c|c|}    
		\hline 
		 i & $\mu_i$ (MPa)& $\alpha_i$ & D$_i$ \\
		\hline 
		1  &  14.9637655 &  1.14899635 &  2.603E-03 \\
		\hline 
		 2    &    212.735555    &    11.8906340 & 0  \\
		\hline 
		3 & -106.383163  &    -5.94506580 & 0 \\
		\hline
	\end{tabular} 
\end{table}
\end{minipage}
  \end{tabular}
\end{table}

{\bf Glass Microballoons}

The GMBs are modeled as both linear elastic and elastic - perfectly plastic materials within the parametric space considered. The linear elastic properties of GMBs, that is, elastic modulus $E$ and Poisson's ratio $\nu$, are considered to be 60 GPa and 0.31 \cite{JAYAVARDHAN2018}, respectively. A range of plastic yield strengths are considered from 100 MPa to 10,000 MPa in the elastic-perfectly plastic model after the linear elastic regime to model crushing effects in microballoons. The formation of multiple plastic hinges drive the collapse of microballoons.

% plastic hinge formation in GMB particles that model 

{\bf GMB/HDPE interface}

The interfacial behavior between the GMBs and HDPE matrix is considered as a parameter in this study. Both perfect and imperfect bonding are considered for investigating the influence of interfacial bonding on the load transfer and consequently densification stress and energy absorption. Frictional contact between the microballoons and the matrix is considered, where the coefficient of friction $\mu_f$ of 1 has effects similar to perfect bonding, that is, minimal to no slippage at the interface. This was verified by comparing the compressive stress-strain responses of models with $\mu_f = 1$ and those models where the boundaries between the microballoons and the matrix region were merged, thereby, allowing no debonding or slippage. Imperfect bonding is modeled with $\mu_f =$ 0.1 and 0.01 between the microballoons and the matrix region, which represents intermediate and no bonding, respectively. Essentially, lower coefficients of friction are used for modeling higher extent of imperfect bonding that can allow separation between the microballoons and the matrix region.

\subsection{Parametric Space}\label{parametricSpace}
Table \ref{tab:parametric} shows the types and range of parameters considered in this study. Four different GMB volume fractions V$_{mb}$ are considered, which includes 0\% (pure HDPE resin), 20\%, 40\% and 60\% by volume of GMBs in the model domain. Varying extents of interfacial bonding between GMBs and HDPE resin are modeled by considering coefficient of friction values of $\mu_f$ = 1 (perfect bonding), 0.1 (intermediate bonding) and 0.01 (no bonding). Compressive stress-strain responses are first determined with linear elastic properties of GMBs. This is followed by considering elastic - perfectly plastic properties for the GMBs to model their crushing under compressive loading. The plastic yield strengths ($\sigma_p$) considered are 100, 1,000 and 10,000 MPa. In each model, all GMBs are considered to have the same strength values, and a distribution is not considered here. It is worth noting that crushing can initiate at some weaker GMBs first followed by stronger ones. Hence, this is a limitation of our model. Two wall thicknesses ($t$) of 1.08 $\mu m$ \cite{JAYAVARDHAN2018} and 2.16 $\mu m$ (double that of the first) are considered for the GMBs. The outer radius of the GMBs is maintained constant at 22.5 $\mu m$ and the wall thickness changes towards the interior of the hollow particle. This ensures that the matrix volume fraction is maintained, while the GMB thickness is varied.

\begin{table}[H]
\caption {Parameters considered in this study}
\label{tab:parametric}
\centering
\begin{tabular}{|c|c|}
    \hline 
	 {\bf Parameter} & {\bf Parameter Range}  \\   
    \hline 
	 GMB Volume Fraction (V$_{mb}$ \%) & 0, 20, 40, 60 \\
	 \hline
	 Coefficient of Friction at GMB/HDPE Interface ($\mu_f$) &  1.0, 0.1, 0.01\\
	\hline 
	 GMB Material Type & Linear Elastic, Perfectly Plastic  \\
	\hline 
	GMB Plastic Yield Strength ($\sigma_p$ MPa) &  10$^2$, 10$^3$, 10$^4$ \\
	\hline 
	GMB Wall Thickness ($t$ $\mu m$)    &   1.08 , 2.16       \\
	\hline
\end{tabular} 
\end{table}

\subsection{Multiple Linear Regression Analysis}

Multiple linear regression analysis explores the relationships between several independent variables ($x_1,x_2,...,x_n$) and a targeted dependent variable ($y$) through fitting a linear equation as shown in Equation~\ref{eqn:regression}. Coefficients of this linear equation ($a_0,a_1,...,a_n$) describe how each independent variable controls the value of the target variable, 
\begin{equation}
    y=a_0+a_1x_1+a_2x_2+a_3x_3+a_4x_4+ ... + a_nx_n
\label{eqn:regression}
\end{equation}

Four different parameters shown in Table~\ref{tab:parametric} are used as independent variables (except GMB material type) and four different target variables are considered: densification stress, specific densification stress, densification energy and specific densification energy. Independent variables are re-scaled within the range of [0,1] using min-max scaling defined in Equation~\ref{eqn:norm}. 
\begin{equation}
    x'=\frac{x-\min{x}}{\max{x}-\min{x}}
\label{eqn:norm}
\end{equation}

In this paper, volume fraction (${V_{mb}}$), plastic yield strength ($\sigma_p$) and wall thickness ($t$) are rescaled as $\hat{V}_{mb}$, $\hat{\sigma}_p$ and $\hat{t}$, respectively. Since the minimum value for these three parameters can be zero, we set the minimum value to be zero and maximum value to be the maximum data sample value we obtained. Equation considered for each target variable is shown in Equation~\ref{eqn:regressionEqn}, and details about the \emph{cross terms} will be discussed in section~\ref{sec:multi-reg-result}.
\begin{equation}
\begin{split}
    \textit{densification stress} = a_{01}+a_{11} \hat{t}+a_{21}\hat{\sigma}_p+a_{31}\hat{V}_{mb}+a_{41} \mu_f + \textit{cross terms}\\
    \textit{specific densification stress} = a_{02}+a_{12} \hat{t}+a_{22}\hat{\sigma}_p+a_{32}\hat{V}_{mb}+a_{42} \mu_f + \textit{cross terms}\\
    \textit{densification energy} = a_{03}+a_{13} \hat{t}+a_{23}\hat{\sigma}_p+a_{33}\hat{V}_{mb}+a_{43} \mu_f + \textit{cross terms}\\
    \textit{specific densification energy} = a_{04}+a_{14} \hat{t}+a_{24}\hat{\sigma}_p+a_{34}\hat{V}_{mb}+a_{44} \mu_f  + \textit{cross terms}
    \end{split}
\label{eqn:regressionEqn}
\end{equation}

\section{Results}

\subsection{Compressive Response of GMB/HDPE Syntactic Foams}
\begin{figure}[h!]
	\resizebox{1\textwidth}{!}{
	\begin{tabular}{|c|c|}
	\hline
	Thin Wall & Thick Wall \\
	    \hline 
	{\includegraphics[width=0.5\textwidth]{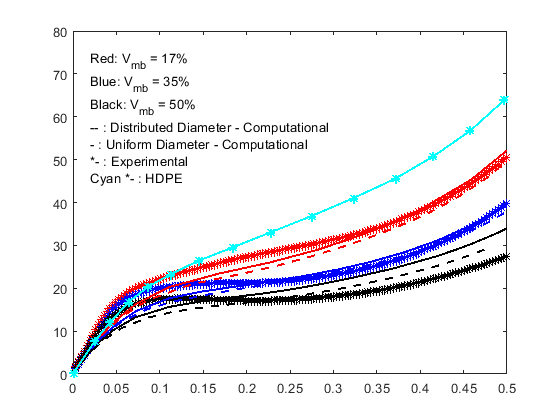}
	(a)}
&
    {\includegraphics[width=0.5\textwidth]{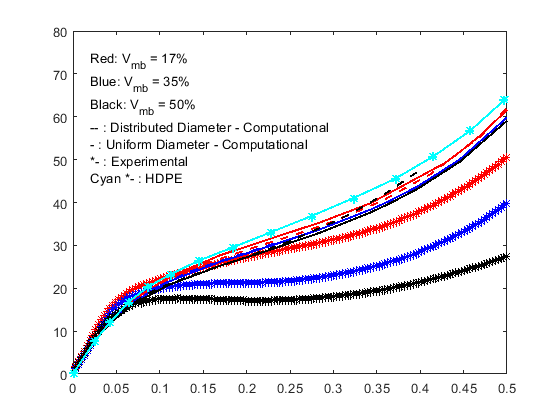} 
    (b)} \\   
	    \hline 
	\end{tabular} }
		\caption{Comparison of experimental and computational compressive stress-strain response of foams with varying GMB volume fraction: (a) Thin wall GMBs; (b) Thick wall GMBs.}\label{stressStrainLE}
\end{figure}

%\subsubsection{Linear Elastic (GMB)/Hyperelastic (HDPE) response}
The hyperelastic response of HDPE was successfully modeled using the experimental compressive stress-strain response obtained from Jayavardhan and Doddamani \mbox{\cite{JAYAVARDHAN2018}} as shown in Figure~\ref{stressStrainLE} (Cyan). The compressive stress-strain response of syntactic foams with varying GMB volume fractions and particle size distributions were modeled first assuming linear elastic properties for the GMBs. The compressive stress-strain responses of syntactic foam models with varying GMB volume fractions, GMB wall thicknesses, and particle size distribution (uniform and polydistribution) are shown in \mbox{Figure~\ref{stressStrainLE}}. GMB volume fractions of 17\%, 35\%, and 50\% are considered in these models to compare against the experimental stress-strain responses from Jayavardhan and Doddamani\mbox{\cite{JAYAVARDHAN2018}}. The wall thickness ($t$) considered in the computational model of the thin wall case (\mbox{Figure~\ref{stressStrainLE}(a)}) matches that of the microballoon wall thickness in the experiments \mbox{\cite{JAYAVARDHAN2018}}.)

A general trend in these stress-strain responses is that increasing GMB volume fraction $V_{mb}$ reduces the densification stresses. Densification stress is defined as the stress corresponding to 0.4 mm/mm compressive strain in this study. We also observe that the GMB size distribution has marginal impact on the stress-strain response compared to that with uniform GMB size. Thus, we will consider uniform GMB size for the remainder of this paper.

As compared to experimental stress-strain graphs, the densification stresses computationally determined for the thin wall case are comparable for lower $V_{mb}$ like 17\% and 35\%, but are higher for higher $V_{mb}$ like in the case of 50\%. This is because, as the $V_{mb}$ increases, their collapse has a higher impact on the densification stresses and thus needs to be captured in the model. Hence, next we considered elastic - perfectly plastic properties for the GMB particles, which will be discussed in the next section. It is worth noting that the difference in the compressive stress-strain responses determined computationally for thick walled GMB models with linear elastic properties have lesser variation with changing V$_{mb}$ as shown in Figure~\ref{stressStrainLE}(b).

\subsection{Densification Stresses and Energy Absorption of GMB/HDPE Syntactic Foams}

\begin{figure}[h!]
	\resizebox{1\textwidth}{!}{
	\begin{tabular}{|c|c|c|}
	\hline
	& Thin Wall & Thick Wall \\
	    \hline 
\rotatebox{90}{No Bonding} & \includegraphics[width=0.5\textwidth]{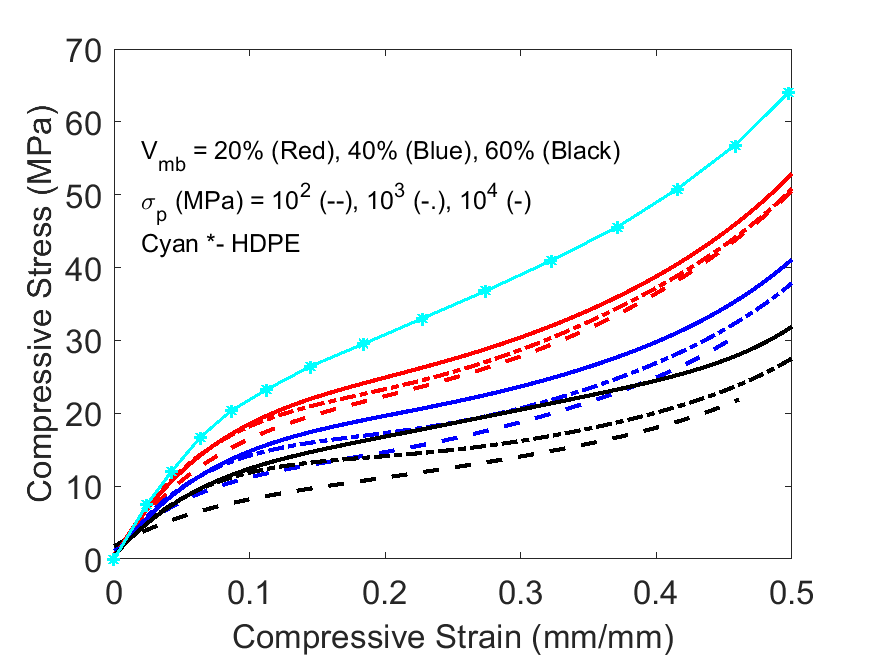} (a)
&
\includegraphics[width=0.5\textwidth]{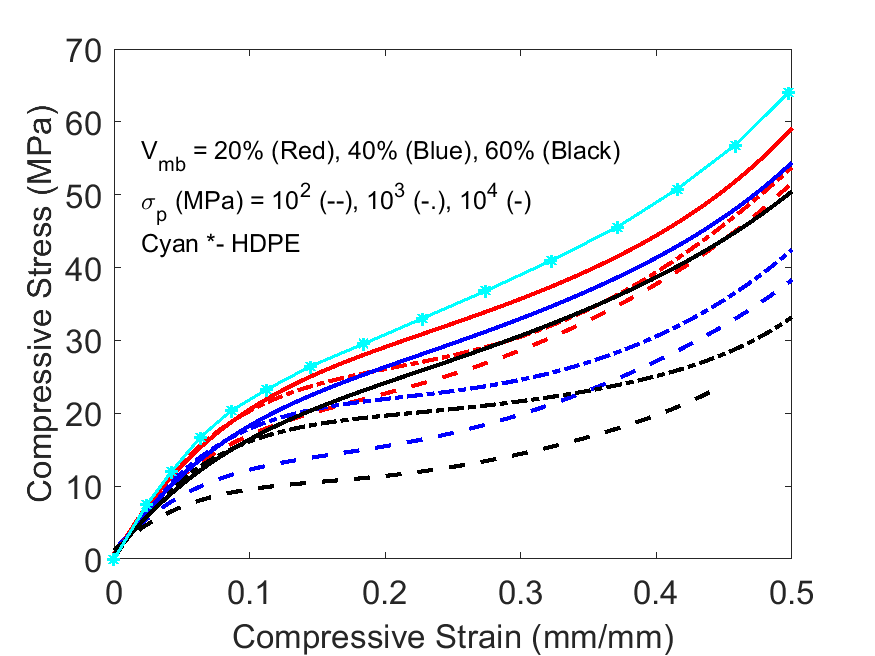} (b) \\   
	    \hline 
\rotatebox{90}{Perfect Bonding} &		 \includegraphics[width=0.5\textwidth]{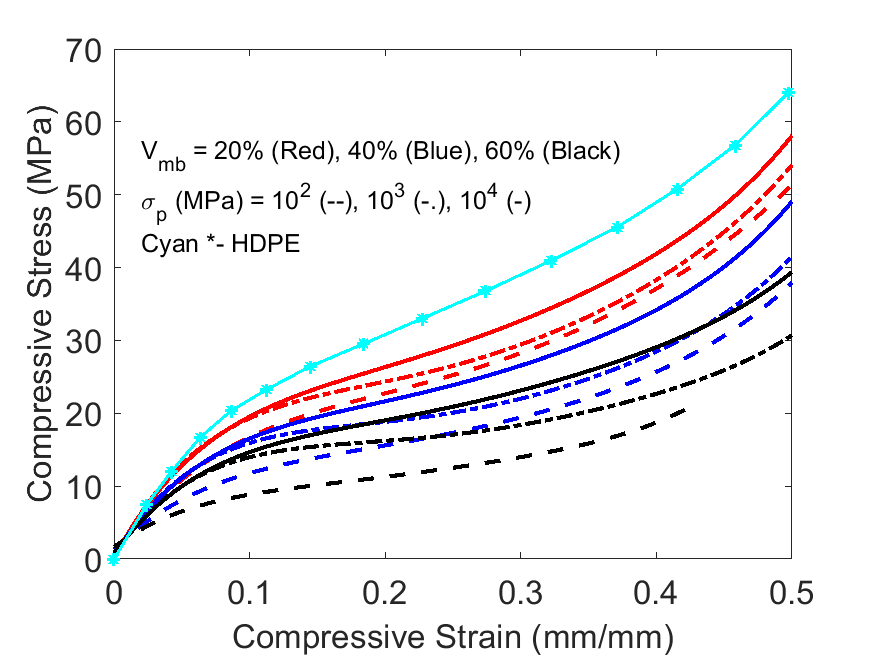} (c)
		 &
		 \includegraphics[width=0.5\textwidth]{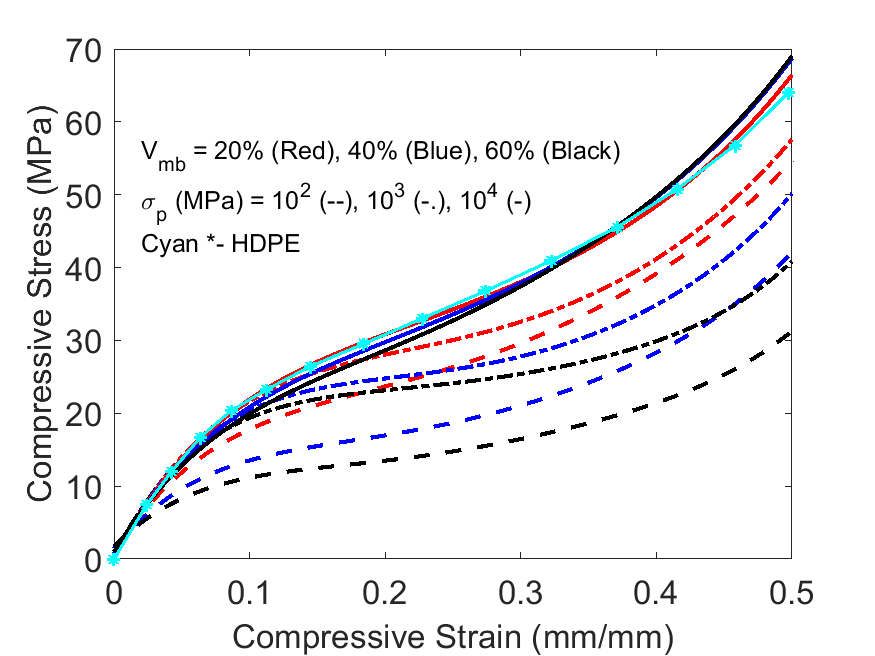} (d) \\
		 \hline
	\end{tabular} }
		\caption{Compressive stress-strain response of foams with varying GMB volume fraction $V_{mb}$. (a) Thin wall thickness and no interfacial bonding; (b) Thick wall thickness and no interfacial bonding; (c) Thin wall thickness and perfect interfacial bonding; (d) Thick wall thickness and perfect interfacial bonding }\label{stressStrainPlas}
\end{figure}

Compressive stress-strain response of GMB/HDPE syntactic foam models with elastic - perfectly plastic properties for the GMBs is presented here. The plastic yield strength ($\sigma_p$) values are considered to be 10$^2$, 10$^3$ and 10$^4$ MPa, which represent different crushing strengths of the GMBs. Figure~\ref{stressStrainPlas} shows the compressive stress-strain graphs for different GMB volume fractions ($V_{mb}$), GMB wall thicknesses ($t$), and interfacial bonding ($\mu_f$) between GMB and HDPE. These are compared against the compressive stress-strain response of pure HDPE.

As expected, increasing the GMB volume fraction $V_{mb}$ decreases the densification stresses. Further, higher GMB $\sigma_p$ increases the densification stresses as higher stresses are required to crush the GMBs. Higher $t$ improves the compressive response, however, special attention should be paid towards weight gain due to an increase in GMB wall thickness. This is addressed in the following sections by presenting maps of weight normalized densification stress and energy absorption values with varying foam parameters. In general, we observe that better interfacial bonding $\mu_f$ results in a slight improvement in the overall compressive response. This is a manifestation of improved load transfer due to good bonding between the microballoons and matrix in the syntactic foams that prevents microballoons from slipping.

% Finally, it is interesting to note that when the interfacial bonding is perfect (i.e. no slippage allowed), syntactic foams with 20, 40 and 60 \% GMB volume fraction and high GMB yield strength (10$^4$ MPa) have overlapping responses with that of pure HDPE. 

%is significant when comparing Figure~\ref{stressStrainPlas}(b) and (c), which correspond to different GMB wall thicknesses. That is, (b) has GMBs with thick walls and no interfacial bonding (b), and thin wall thickness with perfect interfacial bonding (c). These responses are very similar, which implies that thick walled GMB particles with poor interfacial bonding have similar effect on the compressive behavior as thin walled GMB particles with excellent interfacial bonding. However, for a specific GMB volume fraction, the weight of (b) is higher than (c), which makes (c) superior for weight saving applications.

\subsubsection{Densification stresses}
\label{sec:densification_stresses}
\begin{figure}[h!]
\centering
	\resizebox{0.75\textwidth}{!}{
	\centering
	\begin{tabular}{|c|c|c|}
	\hline
	& Thin Wall & Thick Wall \\
	    \hline 
\rotatebox{90}{No Bonding} & \includegraphics[width=0.32\textwidth]{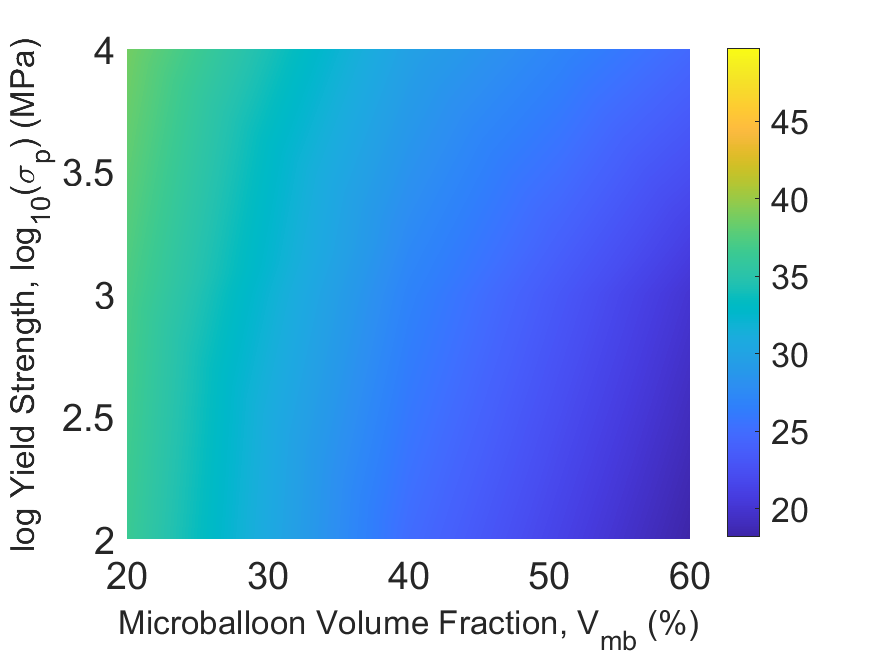} (a)
&
\includegraphics[width=0.32\textwidth]{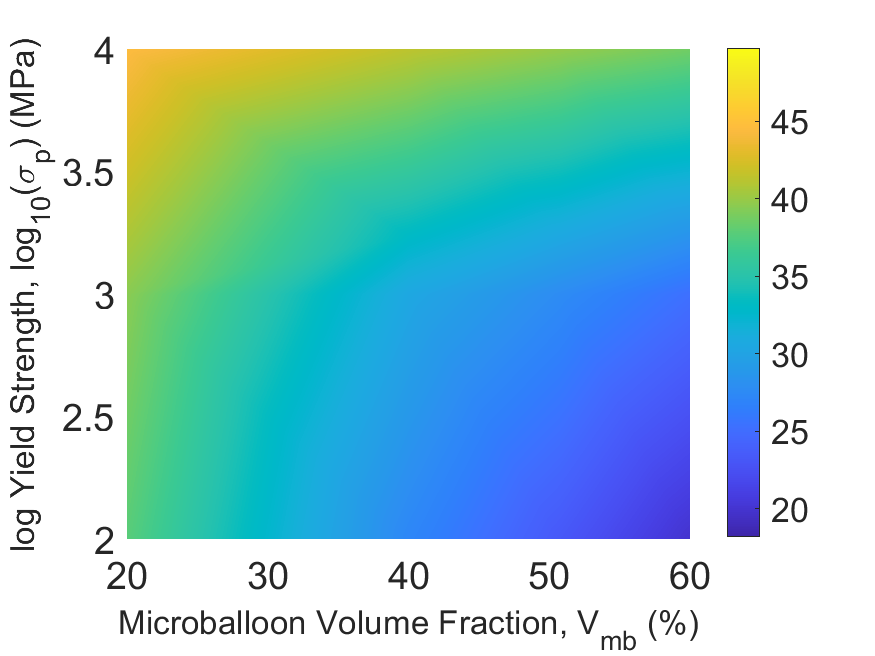} (b)  \\   
	    \hline 
\rotatebox{90}{Perfect Bonding} &		 \includegraphics[width=0.32\textwidth]{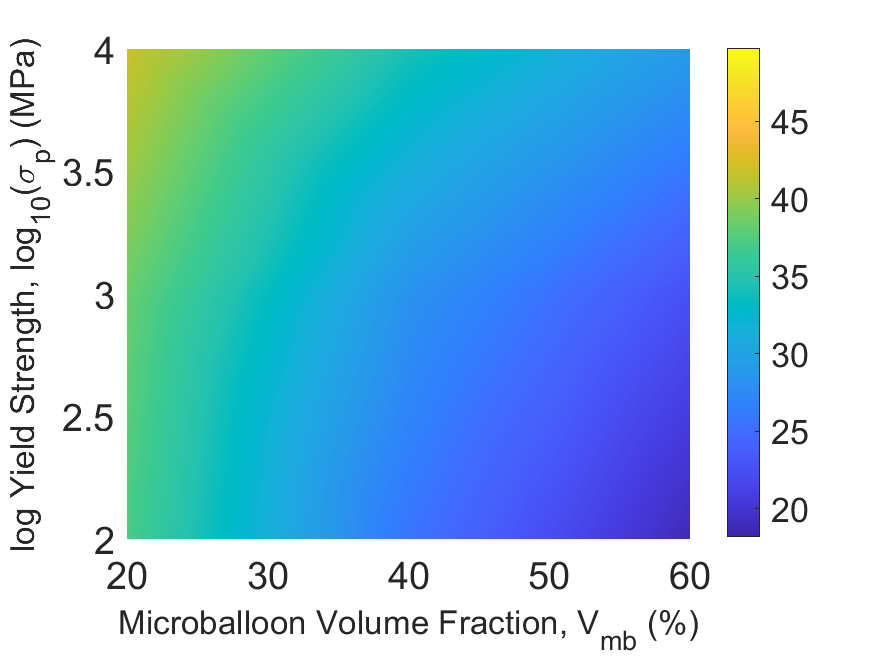} (c)
		 &
\includegraphics[width=0.32\textwidth]{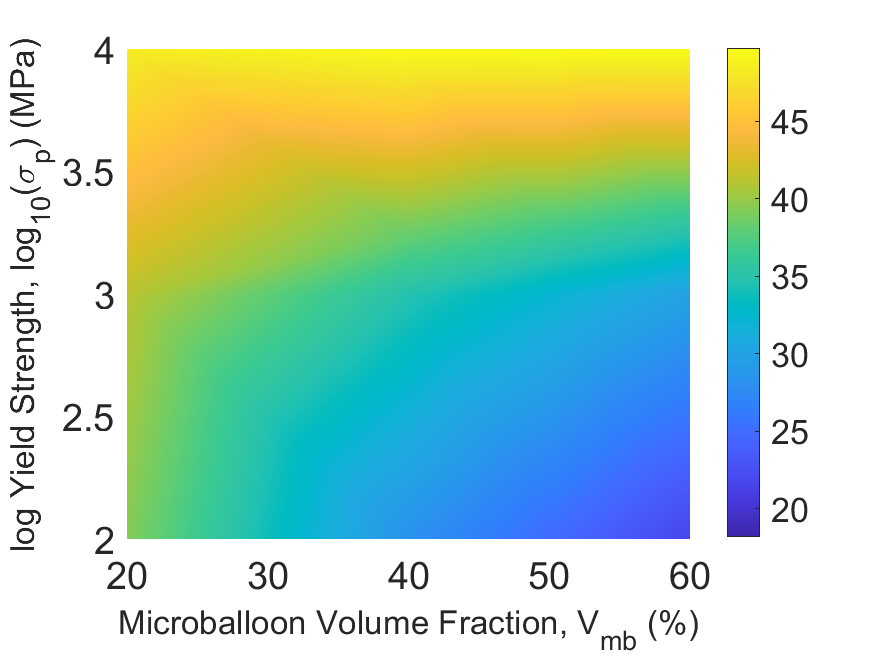} (d) \\
		 \hline
	\end{tabular} 
	}
	\caption{Maps of densification stresses with varying GMB volume fraction $V_{mb}$, plastic yield strength $\sigma_p$, wall thickness and GMB/HDPE interfacial bonding}\label{denseStress}
\end{figure}

Next we compare the densification stresses for the complete parametric space that was investigated within this study. Figure~\ref{denseStress} shows maps of densification stresses which indicate the influence of GMB volume fraction, GMB/HDPE bonding, GMB yield strength, and GMB wall thickness. The densification stresses increase with increasing plastic yield strength for each GMB volume fraction, regardless of interfacial bonding and GMB wall thickness. This is expected as the densification stress depends on the crushing strength of GMB particles.

Although densification stress maps shown in Figure~\ref{denseStress} provide insights into the influence of individual parameters, it is more interesting and valuable to compare the specific densification stresses, that is, weight normalized densification stresses. Note that the weight of syntactic foams decrease with increasing GMB volume fraction, but also increase with increasing wall thickness. 

\begin{figure}[h!]
\centering
	\resizebox{0.75\textwidth}{!}{
	\centering
	\begin{tabular}{|c|c|c|}
	\hline
	& Thin Wall & Thick Wall \\
	    \hline 
\rotatebox{90}{No Bonding} & \includegraphics[width=0.32\textwidth]{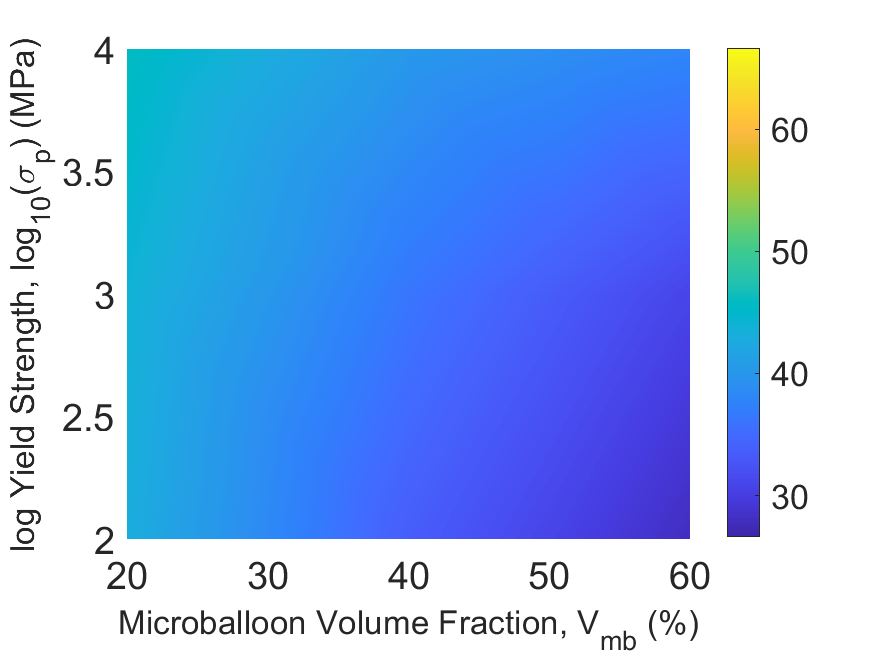} (a)
&
\includegraphics[width=0.32\textwidth]{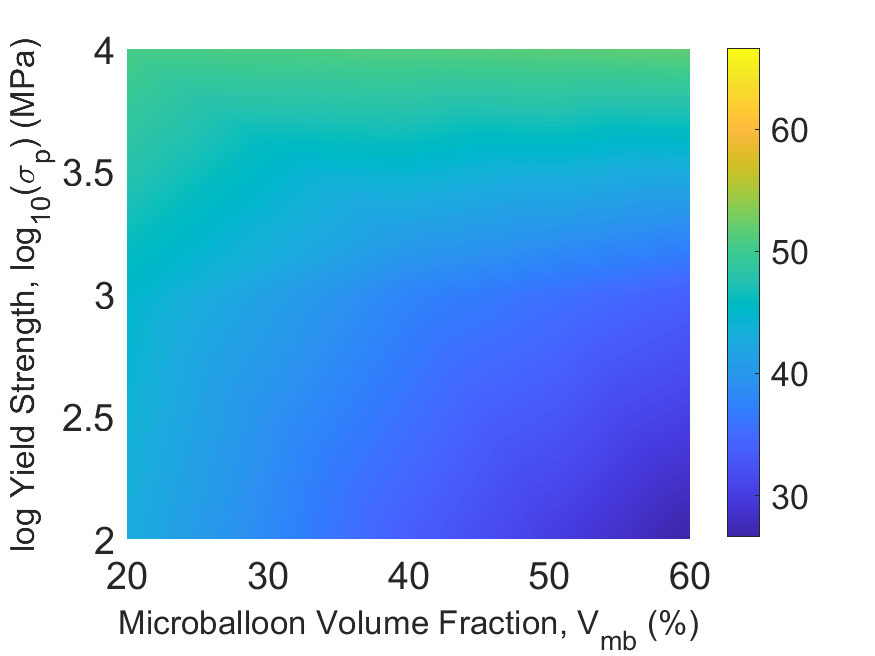} (b)  \\   
	    \hline 
\rotatebox{90}{Perfect Bonding} &		 \includegraphics[width=0.32\textwidth]{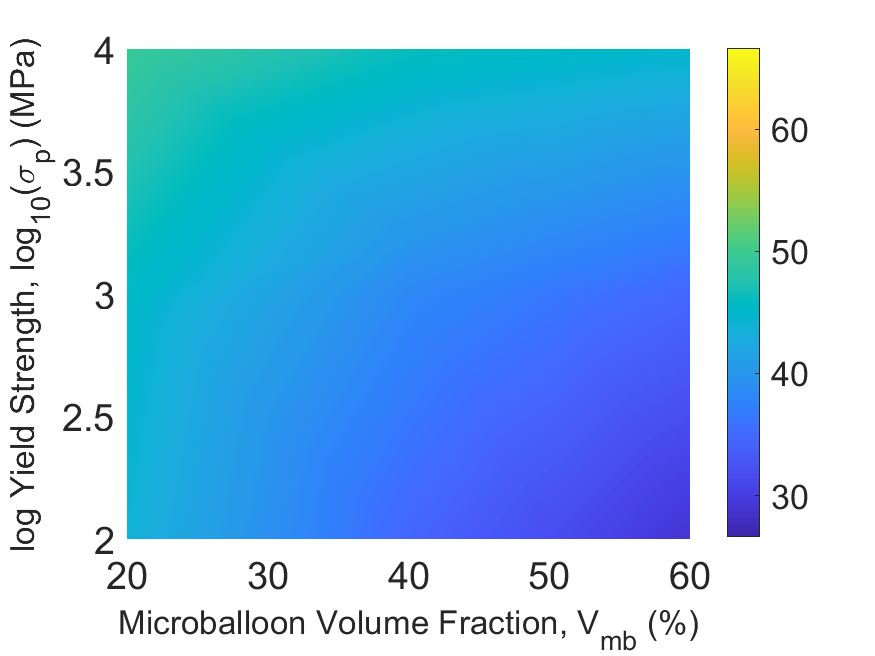} (c)
		 &
\includegraphics[width=0.32\textwidth]{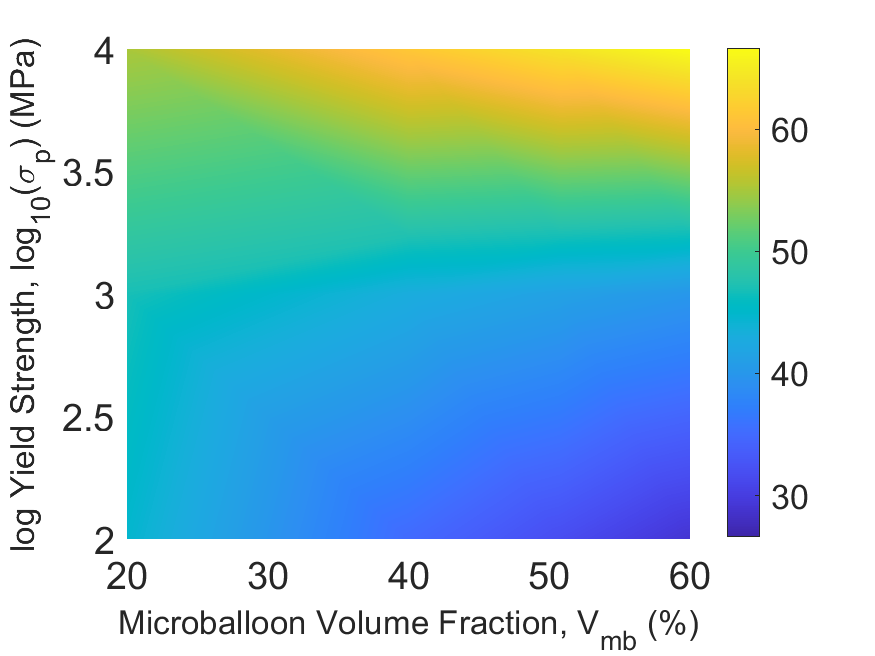} (d) \\
		 \hline
	\end{tabular} 
	}
	\caption{Maps of specific densification stresses with varying GMB volume fraction $V_{mb}$, plastic yield strength $\sigma_p$, wall thickness and GMB/HDPE interfacial bonding}\label{denseStressSpec}
\end{figure}

Figure~\ref{denseStressSpec} shows the maps of specific densification stresses with varying GMB volume fraction, plastic yield strength of GMB particles, GMB wall thickness, and GMB/HPDE interfacial bonding. For syntactic foams with thin walled GMBs (Figure~\ref{denseStressSpec} (a) and (c)), the specific densification stresses decrease with increasing GMB volume fraction for all GMB strength values and the type of GMB/HDPE bonding. Enhanced interfacial bonding does improve the specific densification stresses (Figure~\ref{denseStressSpec}(c)), and its effect is pronounced at higher GMB strengths. For syntactic foams with thick walled GMBs and no bonding at the GMB/HDPE interfaces (Figure~\ref{denseStressSpec}(b)), the trend of specific densification stresses is similar to that of syntactic foams with thin walled GMBs at lower GMB yield strengths. However, a change in this trend is observed at higher GMB yield strengths, and is pronounced in syntactic foams with thick walled GMBs and perfect bonding at GMB/HDPE interfaces (Figure~\ref{denseStressSpec}(d)). That is, the specific densification stresses decrease with increasing GMB volume fraction with low strength GMBs, however, they increase with increasing GMB volume fraction with high strength GMBs. Mathematical description of such trends using regression analysis is shown later in section~\ref{sec:multi-reg-result}.

\begin{figure}[h!]
\centering
	\resizebox{0.75\textwidth}{!}{
	\centering
	\begin{tabular}{|c|c|c|}
	\hline
	& Thin Wall & Thick Wall \\
	    \hline 
\rotatebox{90}{No Bonding} & \includegraphics[width=0.32\textwidth]{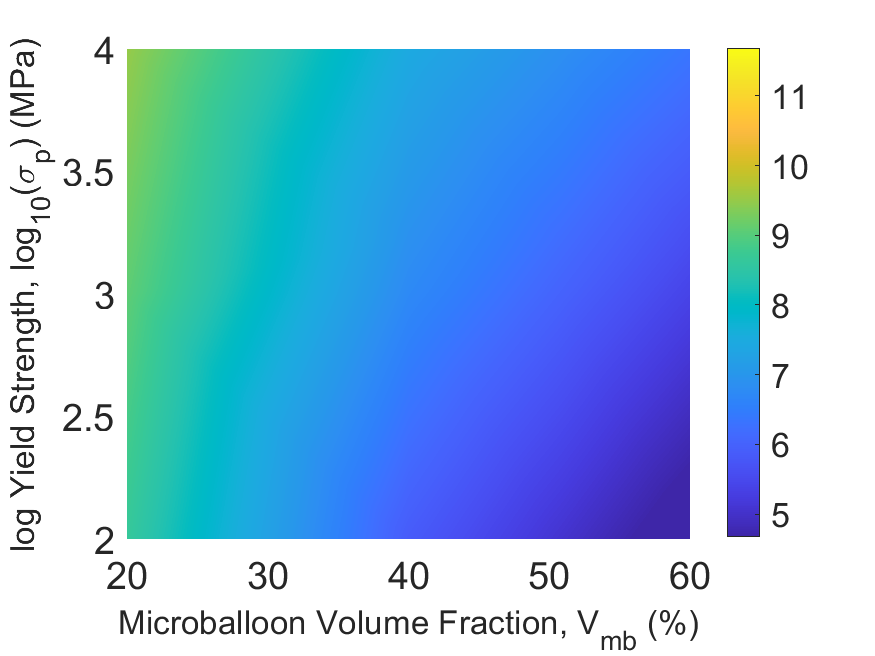} (a)
&
\includegraphics[width=0.32\textwidth]{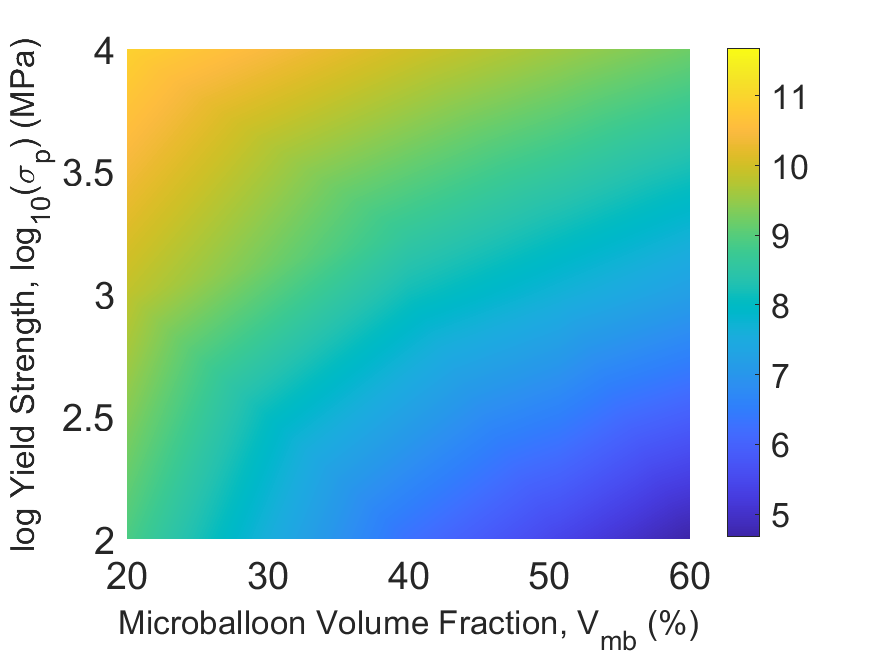} (b) \\   
	    \hline 
\rotatebox{90}{Perfect Bonding} &		 \includegraphics[width=0.32\textwidth]{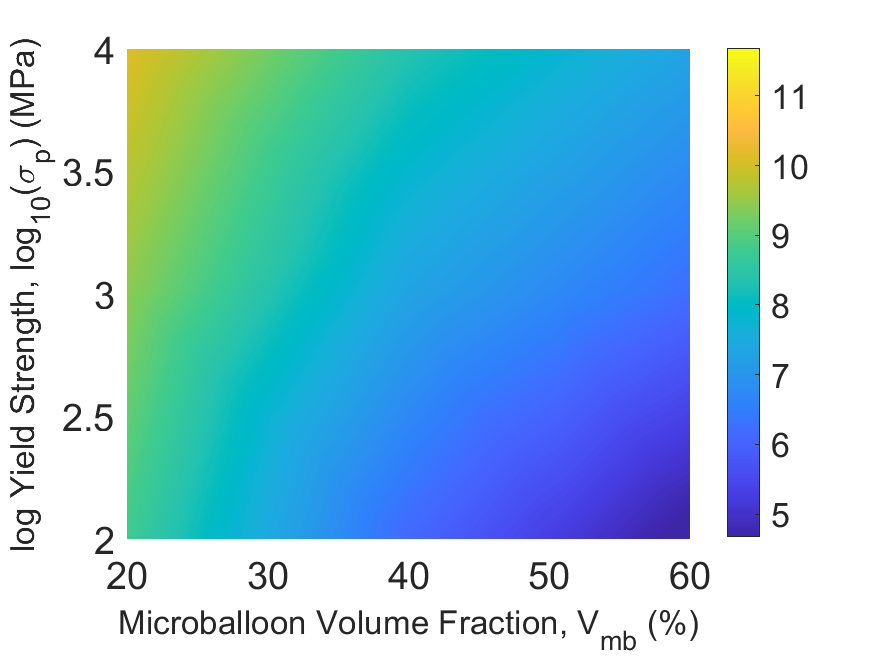} (c)
		 &
\includegraphics[width=0.32\textwidth]{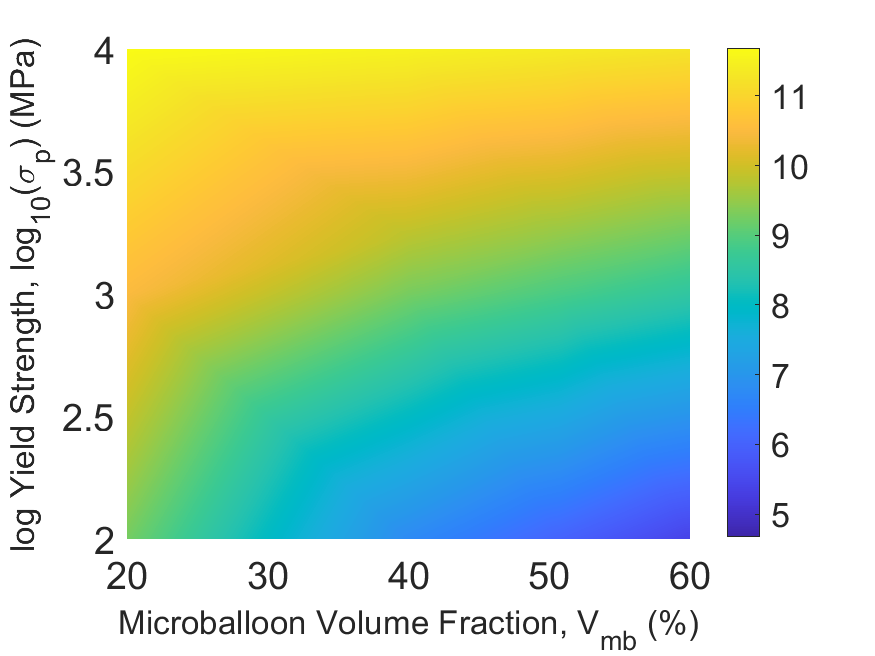} (d) \\
		 \hline
	\end{tabular} 
	}
	\caption{Maps of densification energy with varying GMB volume fraction $V_{mb}$, plastic yield strength $\sigma_p$, wall thickness and GMB/HDPE interfacial bonding}\label{fig:energy}
\end{figure}

\subsubsection{Densification Energy}
\label{sec:energy_absorption}
\begin{figure}[h!]
\centering
	\resizebox{0.75\textwidth}{!}{
	\centering
	\begin{tabular}{|c|c|c|}
	\hline
	& Thin Wall & Thick Wall \\
	    \hline 
\rotatebox{90}{No Bonding} & \includegraphics[width=0.32\textwidth]{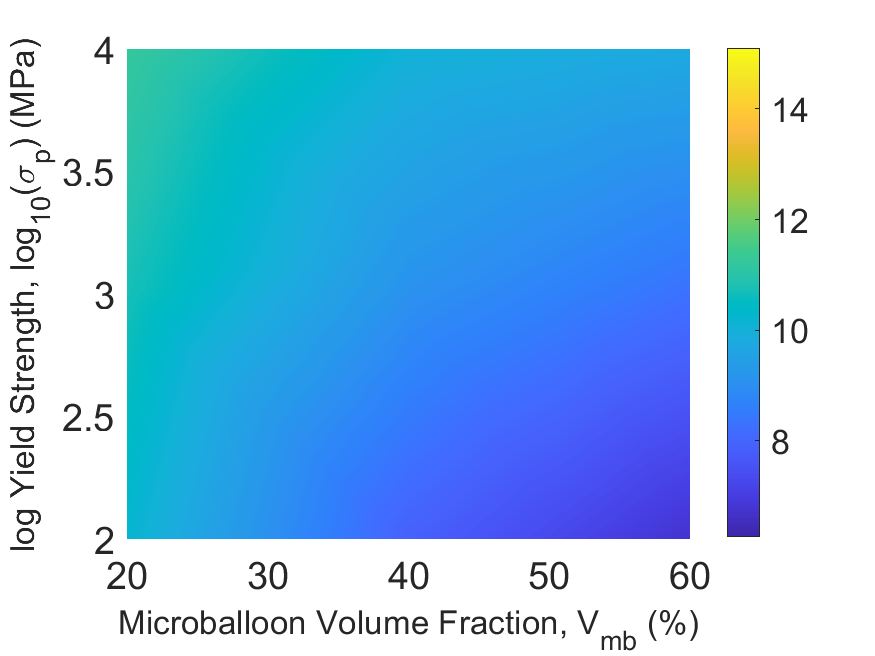} (a)
&
\includegraphics[width=0.32\textwidth]{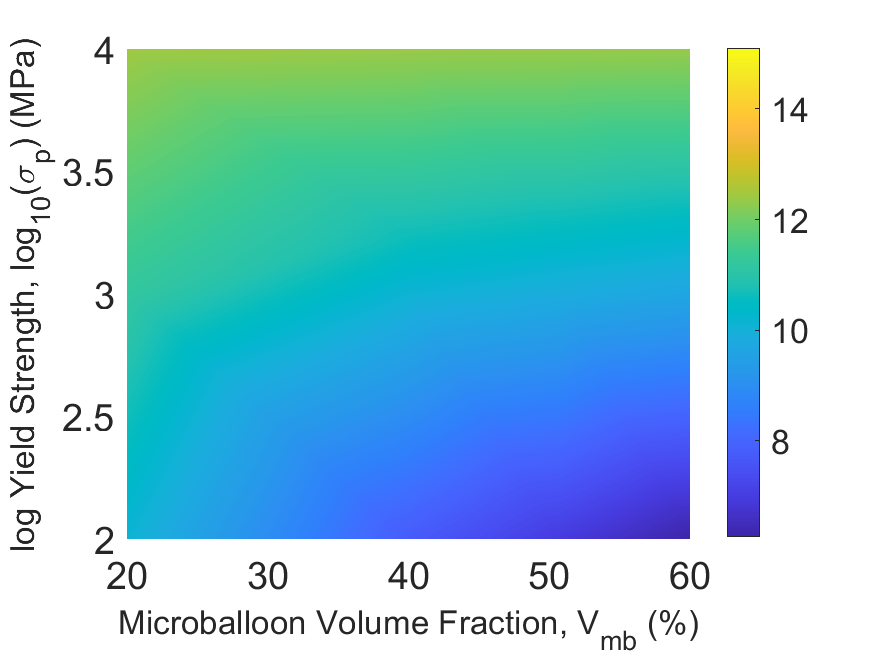} (b)  \\   
	    \hline 
\rotatebox{90}{Perfect Bonding} &		 \includegraphics[width=0.32\textwidth]{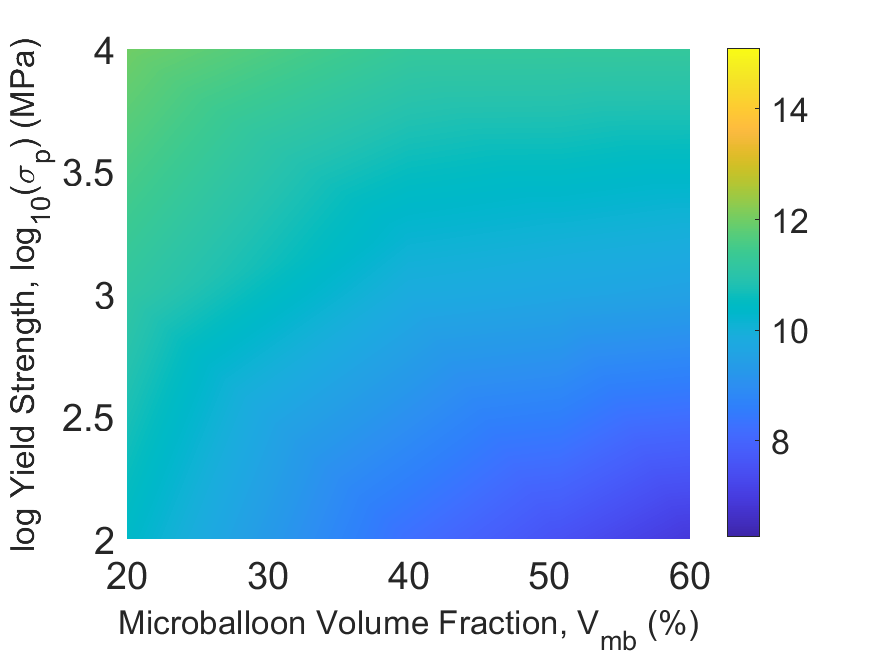} (c)
		 &
\includegraphics[width=0.32\textwidth]{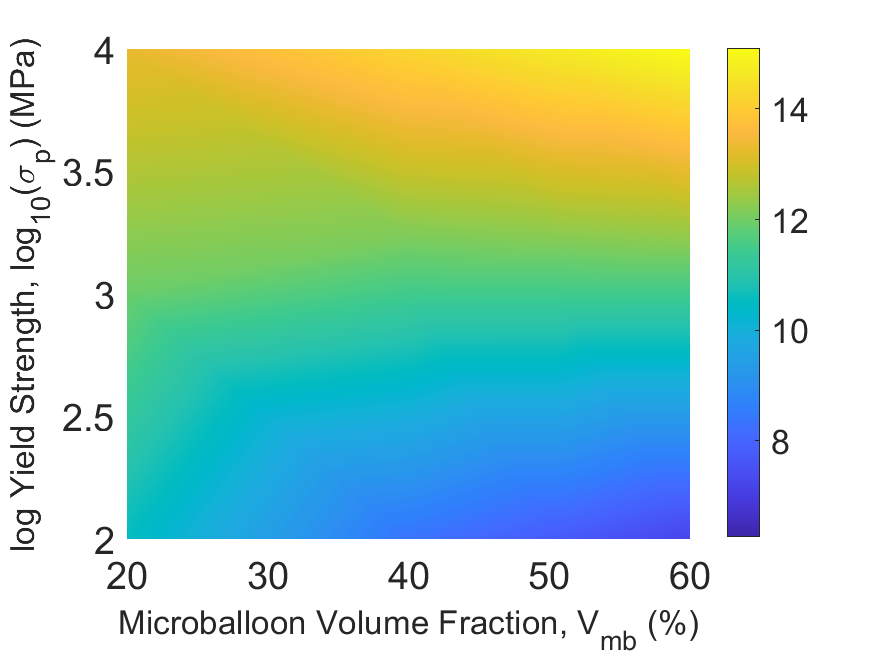} (d) \\
		 \hline
	\end{tabular} 
	}
	\caption{Maps of specific densification energy with varying GMB volume fraction $V_{mb}$, plastic yield strength $\sigma_p$, wall thickness and GMB/HDPE interfacial bonding}\label{fig:energySpec}
\end{figure}

Next we present maps of densification energy similar to that of the densification stresses for all the parameters considered in this study. Densification energy is defined as the area under the compressive stress-strain graphs until a compressive strain of 0.4 mm/mm. Figure~\ref{fig:energy} and Figure~\ref{fig:energySpec} show the maps of densification energy and specific densification energy. The trend in these properties are similar to that of densification stresses and specific densification stresses, respectively.

\subsection{Deformation of GMB/HDPE Syntactic Foams}

\begin{figure}[h!]
\subfigure[Perfect Bonding]{
	\resizebox{0.5\textwidth}{!}{\begin{tabular}{|c|c|c|c|c|}
	\hline
	& \multicolumn{2}{c|}{Low Yield Strength} & \multicolumn{2}{c|}{High Yield Strength} \\
	\hline
	& Thin Wall & Thick Wall & Thin Wall & Thick Wall \\
	    \hline 
\rotatebox{90}{$V_{mb} = 20\%$} & \includegraphics[width=0.08\textwidth]{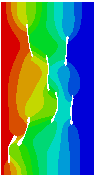}
&
\includegraphics[width=0.08\textwidth]{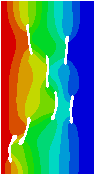} 
&
\includegraphics[width=0.085\textwidth]{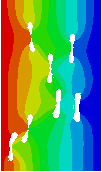}
&
\includegraphics[width=0.082\textwidth]{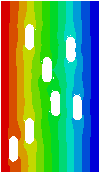} \\
	    \hline 
\rotatebox{90}{$V_{mb} = 60\%$} &		 \includegraphics[width=1.3cm,height=2.3cm]{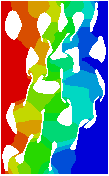}
		 &
\includegraphics[width=1.3cm,height=2.3cm]{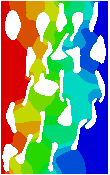} 
&
\includegraphics[width=0.085\textwidth]{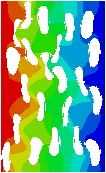} 
&
\includegraphics[width=0.08\textwidth]{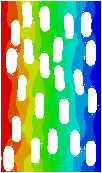} 
\\
		 \hline
	\end{tabular}} 
	}
\subfigure[No Bonding]{
	\resizebox{0.5\textwidth}{!}{\begin{tabular}{|c|c|c|c|c|}
	\hline
	& \multicolumn{2}{c|}{Low Yield Strength} & \multicolumn{2}{c|}{High Yield Strength} \\
	\hline
	& Thin Wall & Thick Wall & Thin Wall & Thick Wall \\
	    \hline 
\rotatebox{90}{$V_{mb} = 20\%$} & \includegraphics[width=0.083\textwidth]{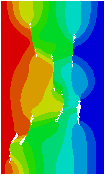}
&
\includegraphics[width=0.082\textwidth]{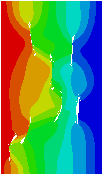} 
&
\includegraphics[width=0.08\textwidth]{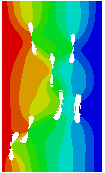}
&
\includegraphics[width=0.08\textwidth]{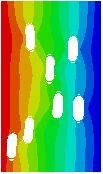} \\
	    \hline 
\rotatebox{90}{$V_{mb} = 60\%$} &		 \includegraphics[width=1.35cm,height=2.4cm]{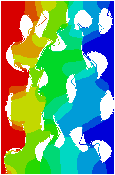}
		 &
\includegraphics[width=1.35cm,height=2.4cm]{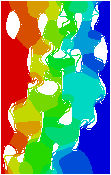} 
&
\includegraphics[width=0.084\textwidth]{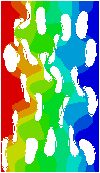} 
&
\includegraphics[width=0.083\textwidth]{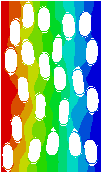} 
\\
		 \hline
	\end{tabular} }
}
\caption{Displacement field along the x-axis at 0.4 mm/mm compressive strain} \label{fig:dispDense}
\end{figure}

Next we visually compare the deformation contour maps of syntactic foams corresponding to a compressive strain of 0.4 mm/mm as shown in Figure~\ref{fig:dispDense}. As compared to Figure~\ref{fig:dispDense}(a), we observe that the walls of the GMB particles are debonded from the matrix (HDPE) in Figure~\ref{fig:dispDense}(b) due to poor or no interfacial bonding. Further, the GMB particles in Figure~\ref{fig:dispDense}(b) are more severely deformed, that is, they have lower stability as compared to those in Figure~\ref{fig:dispDense}(a) when other parameters are held constant. We also observe that GMB particles with higher yield strength and thicker walls have better structural stability against crushing or collapse.

\subsection{Qualitative Comparison with Experiments}
\begin{figure}[h!]
\centering
	\begin{tabular}{|c|c|c|c|}
	\hline
     & $V_{mb} = 20\%$ & $V_{mb} = 40\%$ & $V_{mb} = 60\%$ \\
	    \hline 
\rotatebox{90}{Low Yield Strength} & (a) \includegraphics[width=0.2\textwidth]{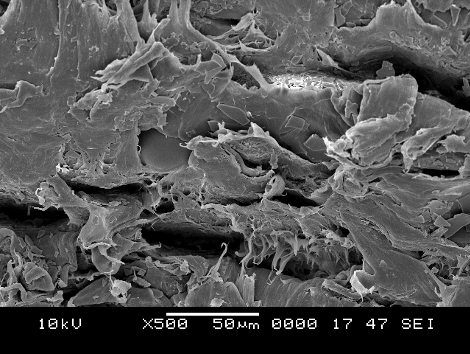}
&
(b) \includegraphics[width=0.2\textwidth]{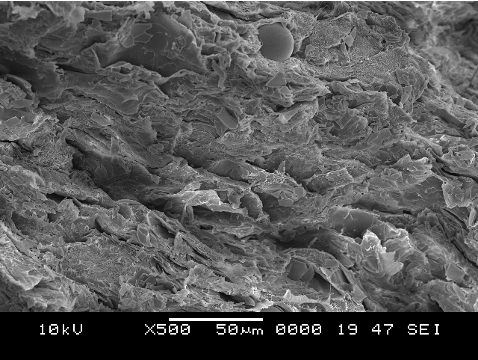} 
&
(c) \includegraphics[width=0.2\textwidth]{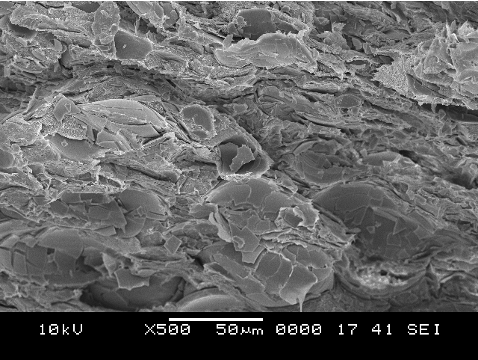}
 \\
	    \hline 
\rotatebox{90}{High Yield Strength} &	(d) \includegraphics[width=0.2\textwidth]{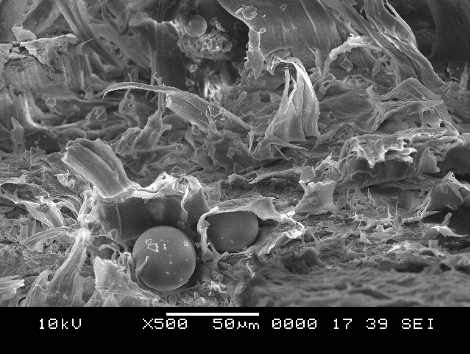}
		 &
(e) \includegraphics[width=0.2\textwidth]{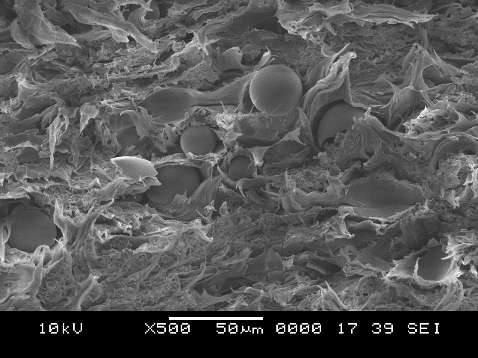} 
&
(f) \includegraphics[width=0.2\textwidth]{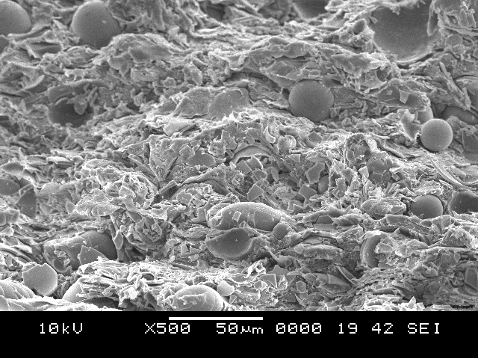} 
\\
		 \hline
	\end{tabular}
\caption{Fractography images of syntactic foams with varying yield strength and GMB volume fractions} \label{fig:sem}
\end{figure}

Scanning Electron Microscopy (SEM) images of failed surfaces of syntactic foam samples upon quasi-static compression \cite{JAYAVARDHAN2018} are shown in Figure~\ref{fig:sem}. The columns correspond to the images of samples with GMB volume fractions of 17\%, 35\%, and 50\%. The rows correspond to microballoons with different collapse pressures, which is linked to the compressive yield strength of the GMB particles. We notice that as the collapse pressure and correspondingly the compressive strength of GMBs increases, more GMBs have survived at all volume fractions. This indicates that the yield strength has a significant impact on the crushing behavior and corresponding densification of syntactic foams.  

\subsection{Identifying the Influence of Individual Parameters} \label{sec:multi-reg-result}

\begin{figure}[h!]
%\centering
\subfigure[]{
    \includegraphics[width=0.5\textwidth]{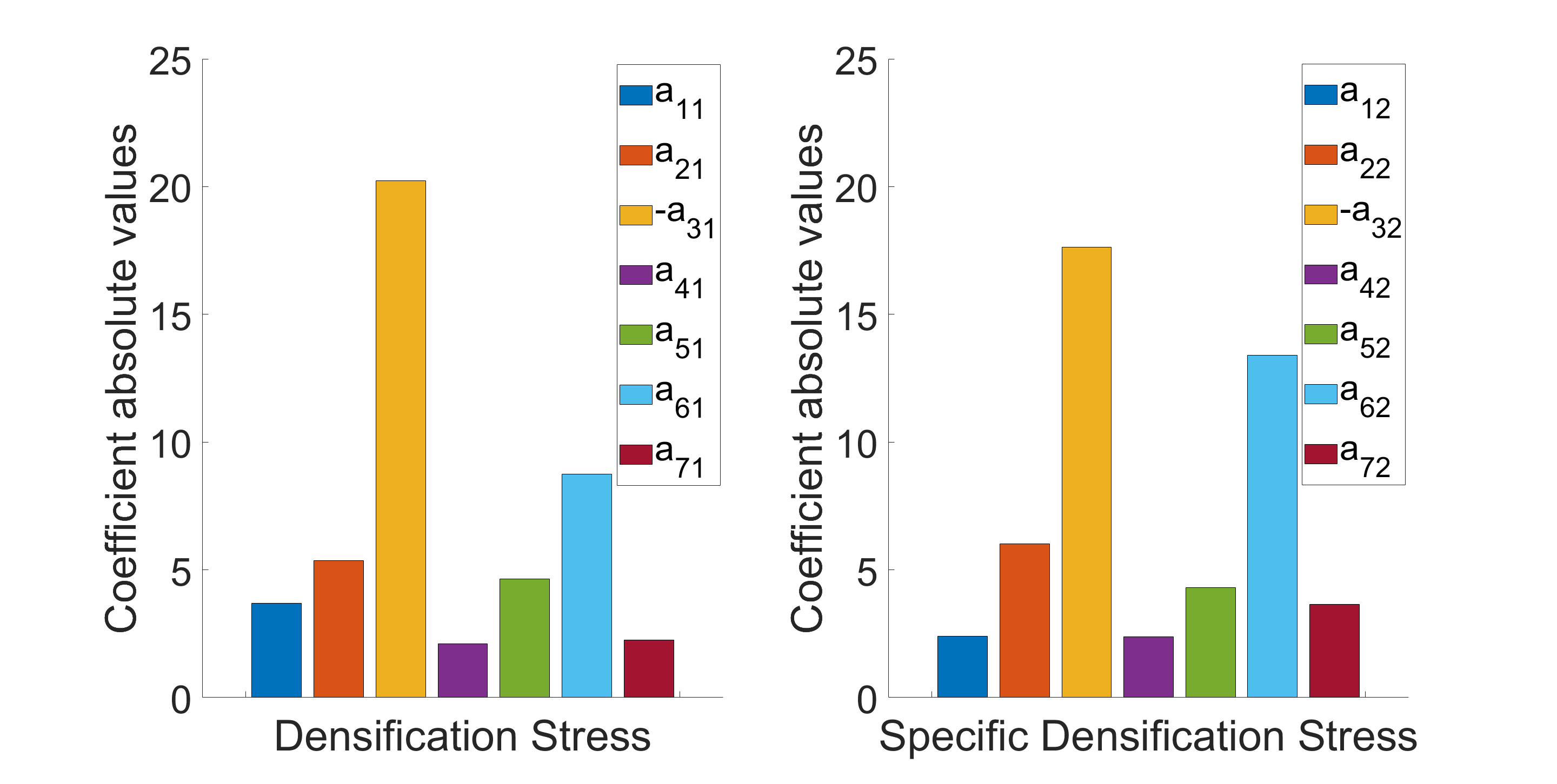}
}
%\centering
\subfigure[]{
    \includegraphics[width=0.5\textwidth]{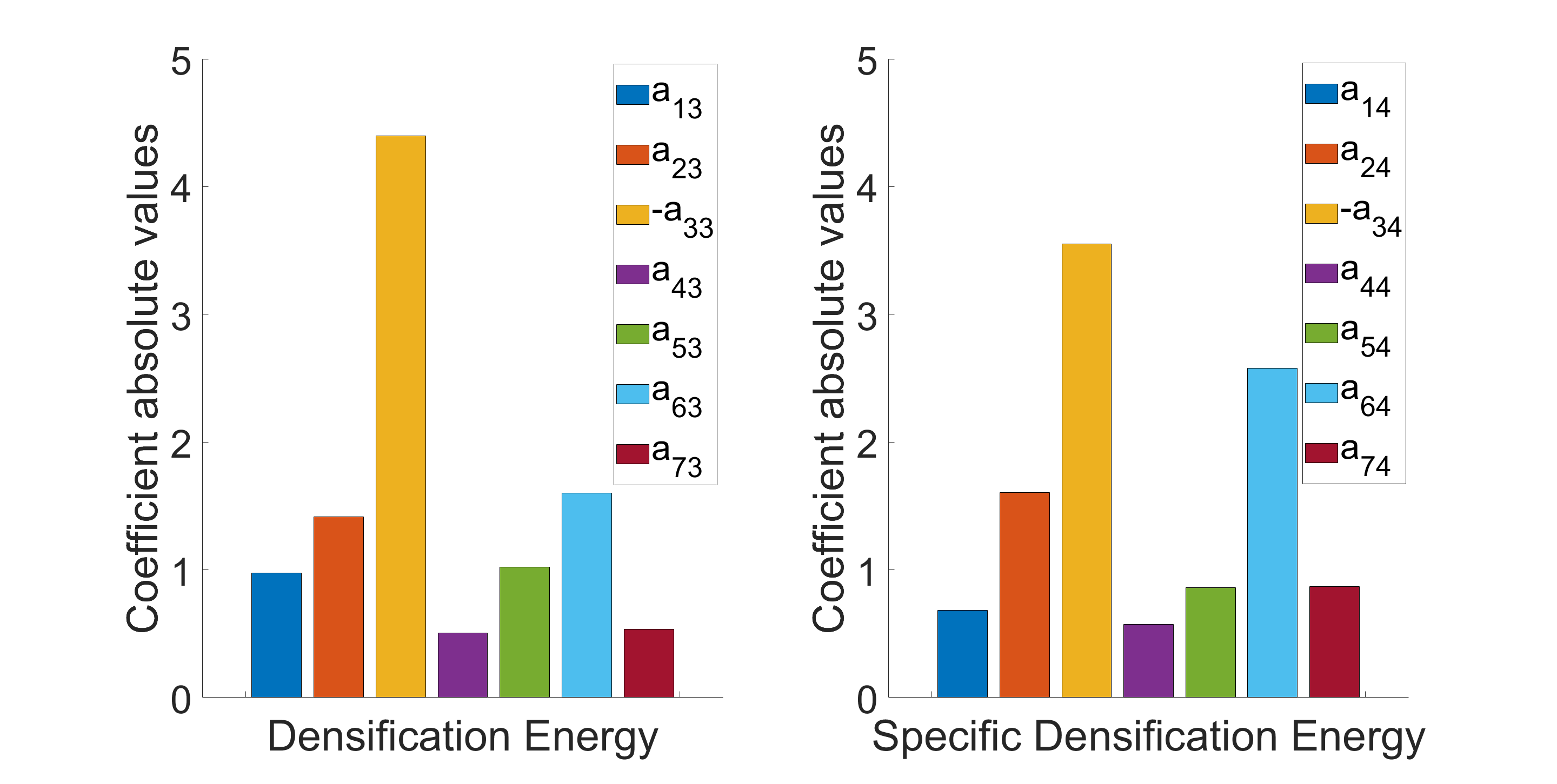}
}
\caption{Coefficients from multiple linear regression analysis}
\label{img:stat_coefficient}
\end{figure}

Although insightful, the maps of densification stresses, energy, and their specific values shown in Figures~\ref{denseStress} to \ref{fig:energySpec} do not provide clear distinction about particular parameters that have the highest or the lowest influence on these properties. Hence, we present here the results from a multiple linear regression analysis that will aid us in identifying the influence of individual parameters considered in this study. We first determine the form of the linear regression models for the four targeted variables (densification stress, specific densification stress, densification energy, and specific densification energy) as shown in Equations~\ref{eqn:regressionEqn}. In addition to wall thickness ($t$), yield stress ($\sigma_p$), microballoon volume fraction ($V_{mb}$), and coefficient of friction ($\mu_f$), Figures~\ref{denseStressSpec} and \ref{fig:energySpec} reveal that cross terms related to ${V_{mb}}$ can change the trends in specific densification stress and specific energy absorption. For instance, in Figure~\ref{denseStressSpec}(d), specific densification stress reduces with increasing volume fraction at yield strength of 100 MPa. However, this property increases with increasing volume fraction at yield strength of 10,000 MPa. This indicates the existence of cross term $\hat{\sigma}_p\hat{V}_{mb}$ that needs to be considered in our regression model. Similarly, comparisons between Figure~\ref{denseStressSpec}(b) and Figure~\ref{denseStressSpec}(d) as well as Figure~\ref{denseStressSpec}(c) and Figure~\ref{denseStressSpec}(d) suggest that cross terms $\hat{t}\hat{V}_{mb}$ and $\mu_f\hat{V}_{mb}$ may have a impact on the regression model. Thus, the expanded expression for Equation~\ref{eqn:regressionEqn} is shown in Equation~\ref{eqn:regressionEqnCross}.
{\small
\begin{equation}
\begin{split}
%\setupformulas[align=right,leftmargin=3em]
    \textit{densification stress} = a_{01}+a_{11} \hat{t}+a_{21}\hat{\sigma}_p+a_{31}\hat{V}_{mb}+a_{41} \mu_f + a_{51}\hat{t}\hat{V}_{mb}+a_{61}\hat{\sigma}_p\hat{V}_{mb}+a_{71}\mu_f\hat{V}_{mb} \\
    \textit{specific densification stress} = a_{02}+a_{12} \hat{t}+a_{22}\hat{\sigma}_p+a_{32}\hat{V}_{mb}+a_{42} \mu_f + a_{52}\hat{t}\hat{V}_{mb}+a_{62}\hat{\sigma}_p\hat{V}_{mb}+a_{72}\mu_f\hat{V}_{mb}\\
    \textit{densification energy} = a_{03}+a_{13} \hat{t}+a_{23}\hat{\sigma}_p+a_{33}\hat{V}_{mb}+a_{43} \mu_f + a_{53}\hat{t}\hat{V}_{mb}+a_{63}\hat{\sigma}_p\hat{V}_{mb}+a_{73}\mu_f\hat{V}_{mb}\\
    \textit{specific densification energy} = a_{04}+a_{14} \hat{t}+a_{24}\hat{\sigma}_p+a_{34}\hat{V}_{mb}+a_{44} \mu_f  + a_{54}\hat{t}\hat{V}_{mb}+a_{64}\hat{\sigma}_p\hat{V}_{mb}+a_{74}\mu_f\hat{V}_{mb}
\end{split}
\label{eqn:regressionEqnCross}
%\end{align*}
\end{equation}
%\stopformula
}
We obtained the coefficients for each dependent variable by performing multiple linear regression analysis in Matlab and Python. The coefficients are shown in Figure~\ref{img:stat_coefficient} and listed in Table~\ref{tab:coeff_reg_analysis}. Note that the sign of $a_{3i}$ is negative. The values of coefficients are further validated with bootstrapping method, which expands the data set and delivers more stable statistical properties. Relatively high values of cross-term coefficients support the necessity of introducing cross-terms into the regression model. The signs of the first four coefficients ($a_{1i},a_{2i},a_{3i}$ and $a_{4i}$, where $i$=1,4), can be visually verified in Figures~\ref{denseStress} to \ref{fig:energySpec}. For instance, in Figure~\ref{denseStress}(a)-(d), densification stress decreases with increasing $V_{mb}$ (horizontally) while increases with increasing $\sigma_p$ (vertically), indicating that densification stress is negatively correlated with $V_{mb}$ and positively correlated with $\sigma_p$. A comparison between Figure~\ref{denseStress}(a) and (b) proves that the sign for $\hat{t}$ is positive, and that between (b) and (d) proves that the sign for $\mu_f$ is also positive. To summarize how these four independent parameters and cross-terms influence our target variables quantitatively, we group our target variables into 1) densification terms and 2) specific densification terms.

\begin{itemize}

\item \underline{Densification terms: Densification stress and densification energy}

Coefficients of densification stress have trends similar to that of densification energy as discussed previously in section~\ref{sec:energy_absorption}. $\hat{V}_{mb}$ ($a_{31}$ and $a_{33}$) has the highest influence on these targets, followed by the cross term $\hat{\sigma}_p \hat{V}_{mb}$  ($a_{61}$ and $a_{63}$) and $\hat{\sigma_p}$  ($a_{21}$ and $a_{23}$), indicating that microballoon volume fraction and yield strength are the two parameters that have the highest influence over the values of densification stress and energy absorption in the normalized scale. Among all other terms, $\mu_f$ and its corresponding cross term have the least influence. The influence of the coefficients corresponding to variables $\hat{t}, \hat{\sigma}_p, \hat{V}_{mb}$, and $\mu_f$ can be visually inspected from the contour maps shown in Figures~\ref{denseStress} and ~\ref{fig:energy}. For instance, in Figure~\ref{denseStress}, color changes in the horizontal direction (volume fraction) in each plot is more significant than that along the vertical direction (yield strength). A comparison between maps Figure~\ref{denseStress}(a)-(b) and Figure~\ref{denseStress}(a)-(c) indicate that the influence of wall thickness is higher than the extent of bonding. This corroborates our finding from the regression analysis, where the absolute value of coefficient corresponding to volume fraction is the highest followed by yield strength and their corresponding cross term, while the coefficient corresponding to $\mu_f$ has the lowest value. This observation is in-line with that established experimentally by prior researchers \mbox{\cite{Shahapurkar2018,JAYAVARDHAN2018,BharathKumar2016,PATIL2019}} that the influence of interface is lower on the compressive response of syntactic foams compared to that of tensile response.

\item \underline{Specific densification terms: Specific densification stress and specific densification energy}

Coefficients of specific densification stress also have trends similar to that of specific densification energy. $\hat{V}_{mb}$ has the highest influence on these targets, followed by cross term $\hat{\sigma}_p\hat{V}_{mb}$ and variable $\hat{\sigma}_p$. However, when compared to the coefficients in the corresponding densification terms, the relative significance of $\hat{t}$, $\hat{V}_{mb}$ and $\hat{t}\hat{V}_{mb}$ are reduced. An intuitive understanding of this phenomenon is that the specific values are evaluated by per unit mass, where the total mass is a function of $\hat{V}_{mb}$ and $\hat{t}$. Hence, the coefficients of $\hat{t}$ ($a_{12}$ and $a_{14}$), $\hat{V}_{mb}$ ($a_{32}$ and $a_{34}$) and $\hat{t}\hat{V}_{mb}$ ($a_{52}$ and $a_{54}$) are expected to reduce their influence on the specific densification quantities as compared to the coefficients of the densification terms. As a result, the significance of $\hat{\sigma}_p \hat{V}_{mb}$  ($a_{62}$ and $a_{64}$) and $\hat{\sigma_p}$ ($a_{22}$ and $a_{24}$) increased. Thus, we further confirmed our visual observation with our regression model.

\begin{comment}
\item \underline{Regression coefficient change between densification and specific densification terms}

Regression coefficients of densification and specific densification terms can be different as we showed in Figure~\ref{img:stat_coefficient}. One of the question raised up is how does these terms change the regression coefficients. Here we define a Coefficient Ratio $CR = a_{sd}/a_{d}$, where $sd$ denotes specific densification term and $d$ denotes densification term. Figure~\ref{img:coef_ratio} shows a comparison between two terms, where $CR>1$ means the coefficient increases from densification term to specific densification term. From the histogram we can find out that, specific densification term has smaller regression coefficients in GMB wall thickness, GMB volume fraction and their cross-term, for both stress and energy. This indicates that GMB wall thickness and volume fraction have less influence on the values of specific densification terms compared to densification terms. \textcolor{red}{Such change happens because the specific densification terms are normalized over the overall density, while volume fraction and ring thickness are the factors controlling the overall density, thus the corresponding regression coefficients dropped comparing the densification terms}.

\begin{figure}[h!]
\centering
	\includegraphics[width=0.8\textwidth]{figures/coeff_ratio.png}
	\caption{Coefficient Ratio between Densification and  Densification terms}
\label{img:coef_ratio}
\end{figure}
\end{comment}

\end{itemize}

To summarize, GMB volume fraction and crushing strength have higher influence over the densification and specific densification stress and energy terms. These parameters can be controlled during fabrication by choosing the material of GMB for their strength and the amount of particles added to achieve a particular volume fraction. 

\subsection{Design Aspects - Syntactic Foams with Specified Overall Density}

Syntactic foams can be designed with different GMB volume fractions and wall thicknesses. However, it is possible to achieve the same overall density of syntactic foam with multiple different pairs of volume fraction and wall thickness. In this section, we will discuss how the densification and specific densification stresses change for syntactic foams with the same overall density. The conclusions can be extended to densification energy terms as well.

The overall density ($\rho_{o}$) of syntactic foam is a function of the volume fraction and wall thickness as shown in Equation~\ref{eqn:density}. 

\begin{equation}
\begin{aligned}
    \rho_{o}&= \frac{M_o}{\mathrm{Vol}_{o}} = \frac{\rho_{m} \mathrm{Vol}_{m}+\rho_{mb} \mathrm{Vol}_{mb}}{\mathrm{Vol}_{o}} = \rho_m(1-V_{mb}) + \rho_{mb} V_{mb}\frac{r^2 - (r-t)^2}{r^2}
\end{aligned}
\label{eqn:density}
\end{equation}

where, subscript $m$ is for matrix, $mb$ is for microballoon, and $o$ is for the overall syntactic foam. $\rho$ is the density. $V_{mb}$ is the GMB volume fraction in the syntactic foam. M is the mass and Vol is the volume. $r$ is the outer radius of GMBs, and $t$ is the wall thickness.

By rearranging Equation~\ref{eqn:density}, we can derive a relationship between GMB volume fraction and wall thickness as shown in Equation~\ref{eqn:vol_frac}.

\begin{equation}
    V_{mb} = \frac{\rho_o-\rho_m}{\rho_{mb}\frac{r^2-(r-t)^2}{r^2}-\rho_m}
\label{eqn:vol_frac}
\end{equation}

\begin{figure}[h!]
\centering
	\includegraphics[width=0.5\textwidth]{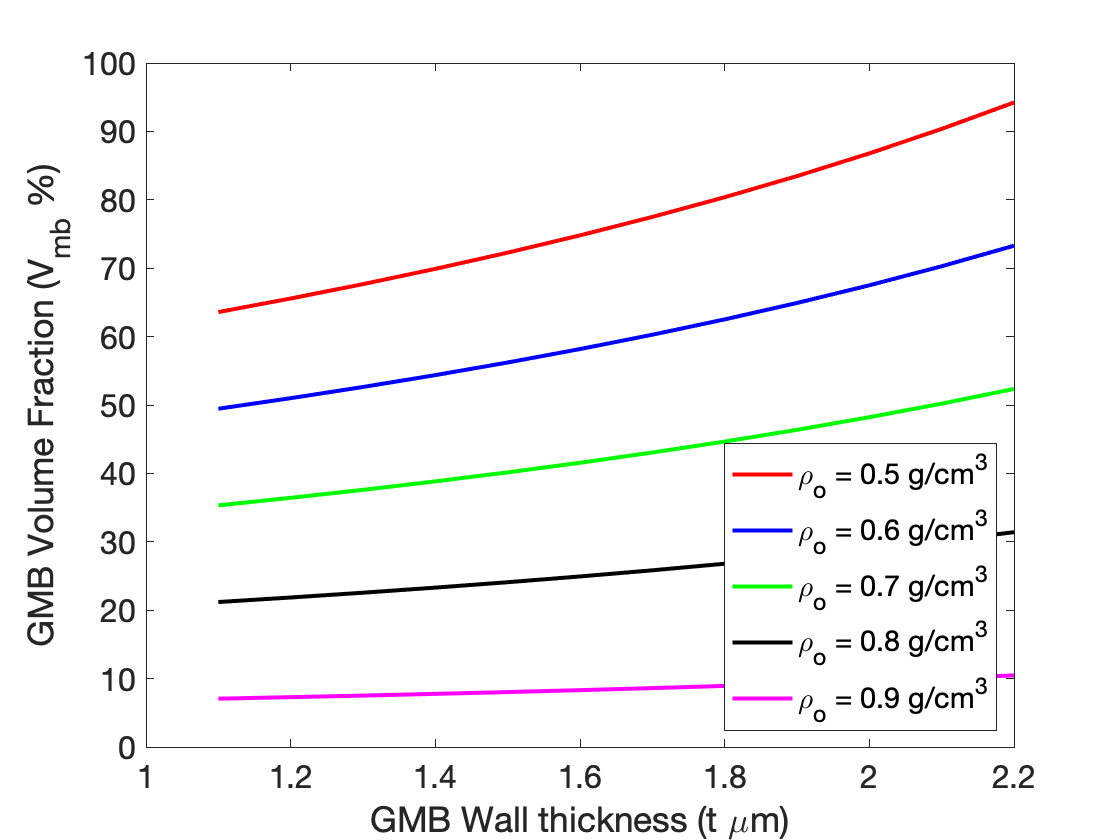}
	\caption{Variation of GMB volume fraction with wall thickness for several specified syntactic foam densities}
\label{img:vf_vs_t}
\end{figure}

We can graph Equation~\ref{eqn:vol_frac} for different values of overall syntactic foam densities as shown in Figure~\ref{img:vf_vs_t}, which is a visualization of the relation between the two parameters for several constant overall densities of syntactic foams. Each curve in this figure is an isoline for density with varying pair of $V_{mb}$ and $t$.

From Figure~\ref{img:vf_vs_t}, we observe that for a fixed overall foam density (that is, a particular curve), the GMB volume fraction increases as the wall thickness increases. That is, syntactic foams with lower volume fractions and thinner GMB walls can have the same overall density as that of foams with higher volume fractions and thicker GMB walls.

To understand how different combinations of GMB wall thicknesses and volume fractions impact the densification and specific densification stresses, we utilize our trained regression model from Section~\ref{sec:multi-reg-result} to estimate theses quantities. For demonstration, we pick the perfect bonding syntactic foam as our test case and a requirement of overall foam density to be $\rho_o=0.66 g/{cm}^3$. Density of HDPE matrix and GMBs are considered to be $\rho_m=0.95 g/{cm}^3$ and $\rho_r=2.54 g/{cm}^3$, respectively. Outer radius of the GMBs is 22.5 $\mu m$. Discrete yield strength ($\sigma_p$) values of 100 MPa, 1000 MPa, 3000 MPa, 6000 MPa, and 10000 MPa are considered for the GMBs. Wall thickness $t$ is allowed to continuously vary from $1.1 \mu m$ to $2.2 \mu m$. Based on Equation~\ref{eqn:vol_frac}, the GMB wall thickness changes will result in $V_{mb}$ to change from 40\% to 60\%. 

\begin{figure}[h!]
%\centering
\subfigure[]{
    \includegraphics[width=0.5\textwidth]{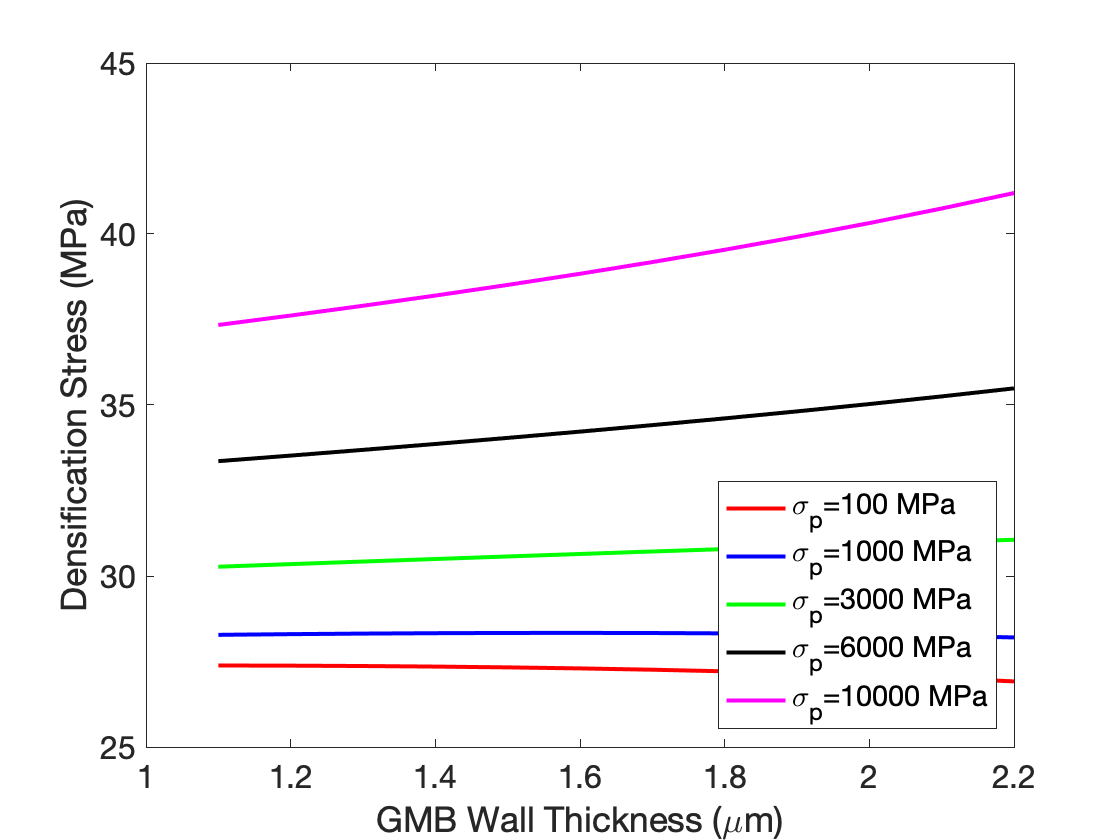}
}
%\centering
\subfigure[]{
    \includegraphics[width=0.5\textwidth]{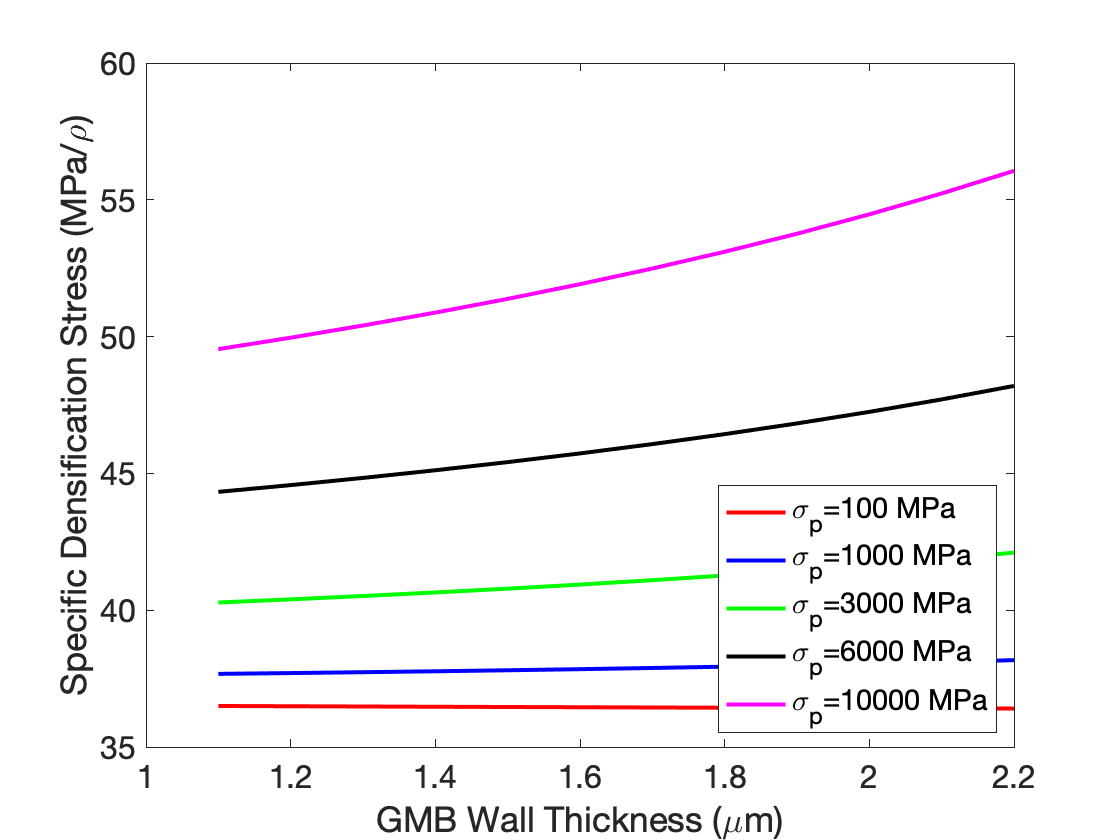}
}
\caption{(a) Densification stress and (b) specific densification stress with varying GMB wall thickness}
\label{img:Den_Stress_T}
\end{figure}

By applying min-max normalization to these new parameters and using our regression model, we obtain the densification stress and specific densification stress plots as a function of wall thickness $t$ as shown in Figure~\ref{img:Den_Stress_T}. From Figure~\ref{img:Den_Stress_T}, we can conclude that for different syntactic foams with the same overall density: (1) both densification stress and specific densification stress will in general increase as the wall thickness increases (and consequently volume fraction increases). (2) higher GMB strength leads to more significant changes in both quantities as the wall thickness increases. Such trends can also be visualized and validated from Figure~\ref{denseStress} and \ref{denseStressSpec}. Thus, if we want higher densification and specific densification stresses, but keeping the syntactic foam density constant, we should choose GMBs with larger wall thickness and at higher volume fractions. Higher strength of the GMBs will further increase these quantities. It is worth noting that the GMBs with thicker walls can have higher crushing strength, adding further to the overall increase in densification quantities.

\section{Conclusion}
In this paper, we presented the mechanisms of densification in syntactic foams with GMBs embedded in HDPE resin. We developed computational models to understand the influence of key syntactic foam parameters including, 
microballoon volume fraction ($V_{mb}$), wall thickness ($t$), yield strength ($\sigma_p$), and the extent of interfacial bonding between microballoons and matrix ($\mu_f$). We elucidated the influence of these parameters on densification stress and energy, and their corresponding specific (weight normalized) values. Further, we developed a multiple regression model to identify the influence of parameters that have higher influence on these properties. Finally, we also performed a case study to explore design aspects of syntactic foams with specified foam density, and the influence of wall thickness and volume fraction on densification properties. Key conclusions from our study are summarized here:

\begin{itemize}
    \item Densification stress and energy increase with increasing plastic yield strength for all GMB volume fractions, regardless of interfacial bonding and GMB wall thickness. This is reflective of the dependence of densification stress and energy on the crushing strength of GMB particles.
    
    \item Although densification properties provide insights into the influence of individual parameters, it is more interesting and valuable to compare the specific densification properties, that is, densification property divided by the weight of the corresponding syntactic foam. The weight of syntactic foams decreases with increasing GMB volume fraction, but increases with increasing wall thickness. 
    
    \item In general, specific densification stress and energy decrease with increasing microballoon volume fraction, and increase with increasing microballoon yield strength for thin walled GMBs. However, we observe a change in this trend for syntactic foams thick walled GMBs at higher yield strength GMBs. Note that wall thickness influences the weight of syntactic foams, but also increase crushing strength of microballoons. Hence, the specific densification stress and energy increase with increasing GMB volume fraction in syntactic foams with high crushing strength and thick walled GMBs.
    
    \item The regression analysis performed in this study corroborated the above-mentioned observations. That is, GMB crushing strength and volume fraction showed higher influence over the values of specific densification stress and specific densification energy, whereas interfacial bonding has the least influence. These parameters can be controlled during fabrication by choosing the material and wall thickness of GMB for their strength, and the amount of particles added to achieve a specified microballoon volume fraction.
    
    \item We further investigated the design aspects for syntactic foams with a specified overall density. For a specified overall syntactic foam density, there exists a one-to-one mapping between the GMB volume fraction and GMB wall thickness. Based on our regression model, we show that a syntactic foam with higher GMB wall thickness and higher volume fraction is preferred over one with lower GMB wall thickness and lower volume fraction, although they could have the same overall foam density. This is because the former case results in higher densification properties. Moreover, higher strength GMBs will further improve these quantities.

\end{itemize}

%\clearpage
\section*{Acknowledgement}
The authors would like to acknowledge the support through the U.S. Office of Naval Research - Young Investigator Program (ONR-YIP) [Grant No.: N00014-19-1-2206] for conducting the research presented here.

% Bibliography
{\footnotesize
\bibliographystyle{unsrt}
\bibliography{syncFoam.bib}
}

\section*{Appendix}
\addcontentsline{toc}{section}{Appendices}
\renewcommand{\thesubsection}{\Alph{subsection}}
\renewcommand{\thetable}{\Alph{subsection}.\arabic{table}}
\setcounter{figure}{0}
\setcounter{table}{0}
\renewcommand{\tablename}{Supplementary Table}

\subsection{Coefficients of Multiple Regression Analysis}
%\begin{comment}
% \usepackage{graphicx}

The coefficients of the variables from the multiple linear regression model shown in Equation~\ref{eqn:regressionEqnCross} are listed below along with the min-max limit values for each variable. Note that the cross terms are calculated after normalizing individual parameters, so we do not apply normalization on cross terms.

\begin{table}[h!]
\centering
\caption{Multiple linear regression analysis}
\resizebox{\textwidth}{!}{%
\begin{tabular}{|c|c|c|c|c|c|}
\hline
 Coefficient of Regression Model & Densification Stress   & Specific Densification Stress   & Densification Energy   & Specific Densification Energy  & Min-Max Value \\ \hline
Intersection      & 34.702        & 41.438       & 8.337 & 9.940  & ---\\ \hline
Wall thickness $\hat{t}$    & 3.686      & 2.411      & 0.973        & 0.684 & [0.0011, 0.0022]\\ \hline
Yield strength $\hat{\sigma_p}$    & 5.366   & 6.020       & 1.416   & 1.605 & [100, 10000]\\ \hline
Volume fraction $\hat{V_{mb}}$    & -20.221  & -17.623     & -4.397  & -3.551   & [20, 60]\\ \hline
Interface bonding $\hat{\mu_f}$ & 2.104      & 2.388    & 0.505      & 0.574 & [0.01, 1]\\ \hline
Cross term $\hat{t}\hat{V_{mb}}$ & 4.635   & 4.314       & 1.020      & 0.863   & --- \\ \hline
Cross term $\hat{\sigma_p}\hat{V_{mb}}$ & 8.755  & 13.392    & 1.602 & 2.578    & --- \\ \hline
Cross term $\hat{\mu_f\hat{V_{mb}}}$ & 2.254     & 3.652     & 0.535  & 0.871   & --- \\ \hline
\end{tabular}%
}
\label{tab:coeff_reg_analysis}
\end{table}
%\end{comment}

\end{document}